\documentclass[preprint]{aastex62}
\usepackage{amsmath,amssymb,CJK,soul}
\usepackage{longtable}
\usepackage{hyperref}
\usepackage{longtable}
\usepackage{threeparttablex} 
\usepackage{lineno}

\newcommand{\pt}{{2024 PT$_5$ }}
\newcommand{\ptns}{{2024 PT$_5$}}

\received{--}
\revised{--}
\accepted{--}
\submitjournal{ApJL}

\shorttitle{Discovery and characterization of minimoon \ptns}
\shortauthors{Bolin et al.}

\begin{document}

\title{The discovery and characterization of minimoon \pt}

\author[0000-0002-4950-6323]{Bryce T. Bolin}
\altaffiliation{These authors contributed equally to this manuscript.}
\affiliation{Eureka Scientific, Oakland, CA 94602, U.S.A.}


\author[0000-0002-7034-148X]{Larry Denneau}
\altaffiliation{These authors contributed equally to this manuscript.}
\affiliation{Institute for Astronomy, University of Hawai`i at M\={a}noa, Honolulu, HI, 96822}

\author[0009-0000-8709-5273]{Laura-May Abron}
\affiliation{Griffith Observatory, Los Angeles, CA 90027}

\author[0000-0001-7830-028X]{Robert Jedicke}
\affiliation{Institute for Astronomy, University of Hawai`i at M\={a}noa, Honolulu, HI, 96822}

\author[0000-0002-9020-5004]{Kristin Chiboucas}
\affil{Gemini Observatory/NSF NOIRLab, Hilo, HI, 96720, USA}

\author[0000-0002-7053-5495]{Carl Ingebretsen}
\affil{William H. Miller III Department of Physics and Astronomy, Johns Hopkins University, Baltimore, MD 21218, USA}

\author[0000-0002-1428-7036]{Brian C. Lemaux}
\affil{Gemini Observatory/NSF NOIRLab, Hilo, HI, 96720, USA}
\affil{Department of Physics and Astronomy, University of California, Davis, One Shields Ave., Davis, CA 95616, USA}




\begin{abstract}
Minimoons are asteroids that become temporarily captured by the Earth-Moon system. We present the discovery of \ptns, a minimoon discovered by the Asteroid Terrestrial-impact Last Alert System (ATLAS) Sutherland telescope on 2024 August 7. The minimoon with heliocentric semi-major axis, $a$$\sim$1.01 au, and perihelion, $q$$\sim$0.99 au, became captured by the Earth-Moon system on 2024 September 29 and left on 2024 November 25 UTC. Visible g, r, i, and Z spectrophotometry was obtained using Gemini North/Gemini Multi-Object Spectrograph (GMOS) on 2024 September 27. The color indices are g-r = 0.58$\pm$0.04, r-i = 0.29$\pm$0.04, i-Z = -0.27$\pm$0.06, and the spectrum best matches lunar rock samples followed by S-complex asteroids. Assuming an albedo of 0.21 and using our measured absolute magnitude of 28.64$\pm$0.04, \pt has a diameter of 5.4$\pm$1.2 m. We also detect variations in the lightcurve of \pt with a 0.28$\pm$0.07 magnitude amplitude and a double-peaked period of $\sim$2600$\pm$500 s. We improve the orbital solution of \pt with our astrometry and estimate the effect of radiation pressure on its deriving an area-to-mass ratio of 7.02$\pm$2.05$\times$10$^{-5}$ m$^2$/kg, implying a density of $\sim$3.9$\pm$2.1 g/cm$^3$, compatible with having a rocky composition. If we assume \pt is from the NEO population, its most likely sources are resonances in the inner Main Belt by comparing its orbit with the NEO population model, though this does not exclude a lunar origin.
\end{abstract}

\section{Introduction}
Evidence from observations of near-Earth objects (NEOs) suggests that the majority originate from the Main Belt asteroids (MBAs) and Jupiter Family Comet (JFC) populations \citep[][]{Granvik2018,Nesvorny2023NEOMOD,Nesvorny2024NEOMOD3}. A small percentage of NEOs can become temporary Earth co-orbitals, located in a 1:1 mean motion resonance with the Earth \citep[][]{Morais2002} where they can become Trojans or quasi-satellites of the Earth for several thousand years \citep[][]{Brasser2004}. A small portion of NEOs, less than $<10^{-7}$ of the total NEO population, can become temporarily captured by the Earth-Moon system's gravity and enter its Hill sphere, becoming a so-called ``minimoon'' \citep[][]{Granvik2013b, Jedicke2018mm}. 

Minimoons are captured at low relative velocities with the Earth-Moon system, less than 1 km/s, where they can be captured on timescales as long as 10s of days to years \citep[][]{Granvik2012, Fedorets2017}. The first known minimoon, 2006 RH$_{120}$, remained in orbit around the Earth for $\sim$400 days \citep[][]{Kwiatkowski2008}, and the second, 2003 CD$_3$ was captured for $\sim$900 days \citep[][]{Naidu2021} before escaping the Earth's Hill sphere. Thus, the capture duration of minimoons represents a small portion of the time they share the Earth's orbit.

While it is assumed that Earth co-orbitals such as minimoons are captured from the NEO population, emerging evidence suggests that some may originate as lunar impact ejecta. Ejecta from lunar impacts can be re-captured onto Earth-similar orbits where they can last for thousands of years \citep[][]{Gladman1995}. Additionally, the observed spectral properties of some minimoons and quasi-satellites such as 2020 CD$_3$ and (469219) Kamo`oalewa resemble some lunar samples \citep[][]{Bolin2020CD3,Sharkey2021}. This evidence has led some to suggest a recent lunar impact as the origin for some Earth co-orbitals \citep[e.g.,][]{Jiao2024}

This paper discusses the discovery of minimoon, \ptns, by the Asteroid Terrestrial-impact Last Alert System (ATLAS) telescope and observations taken at Gemini North. We will use the approach of \citet[][]{Bolin2021LD2,Bolin2022IVO,Bolin2023Dink} to use spectrophotometric observations at different visible wavelengths to constrain its physical properties as a test of whether \pt has an asteroidal or Lunar debris origin. We will also use the astrometry reported from our observations and the observations of others to refine the orbit of \pt, measuring the effect of non-gravitational perturbations on its orbit \citep[e.g.,][]{Micheli2012} and compare its orbit with the NEO population model \citep[][]{Morbidelli2020albedo,Nesvorny2023NEOMOD}.

\section{Observations}
\label{s.obs}
The discovery observations of \pt were taken on 2024 August 7 21:11:57 UTC at the ATLAS Sutherland telescope, Minor Planet Center (MPC) observing code M22, \citep[Fig.~1,][]{Denneau2024PT5}. At the time of the asteroid's discovery, the asteroid was located at right ascension (RA) = 18 57 04.97, dec = -55 56 00.24 and 1.54 lunar distances (0.00397 au) au from the Earth, a heliocentric distance of 1.017 au, at a phase angle of 128.747$^{\circ}$, and had a visible magnitude of $\sim$17. The asteroid was detected in four 30 s ATLAS o-band filter exposures spaced over $\sim$1700 s. The ATLAS o-band filter provides coverage between 560 nm and 820 nm with an effective wavelength of 663 nm \citep[][]{Tonry2018}. The asteroid was moving 13.7$^{\circ}$/d (0.57\arcsec/s) in the northwest direction at the time of discovery, resulting in having a trailed appearance in the ATLAS images (panels a, b, c, and d of Fig~1). The trailed exposures of \pt show little to no curvature over the 1700 s span of the discovery detections with a great circle residual of $\sim$1.2\arcsec.

\begin{figure}
\centering
\includegraphics[scale=0.45]{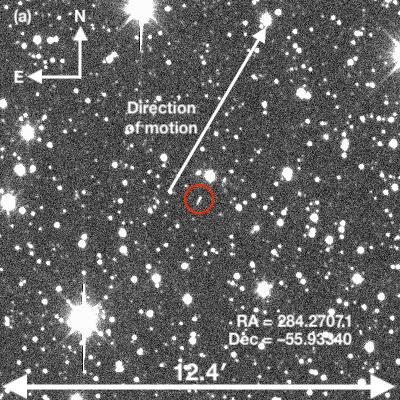}
\includegraphics[scale=0.45]{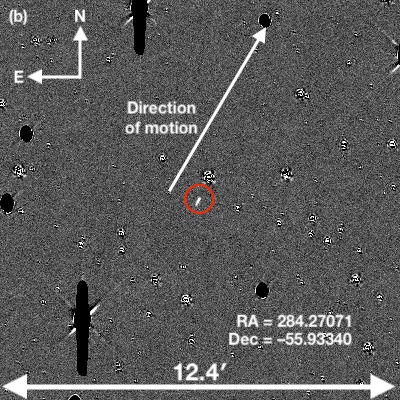}
\includegraphics[scale=0.45]{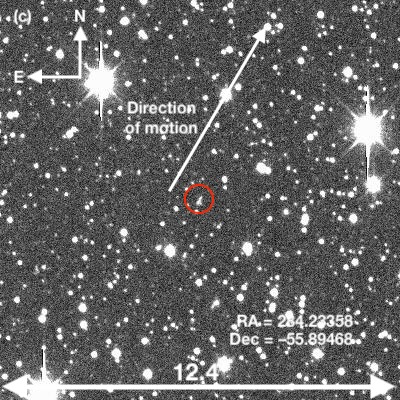}
\includegraphics[scale=0.45]{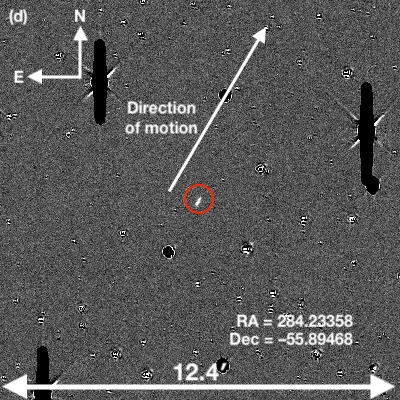}
\caption{\textbf{Panel a:} The first of four o-band ATLAS-Sutherland telescope discovery images of \pt from 2024 August 7 21:11:57 UTC. The asteroid moved at a rate of 34.3 arcminutes per hour (13.7 degrees per day) in the northwest direction. The asteroid makes a $\sim$9 pixel trail in the 30 s ATLAS exposures, indicated by the red circle. \textbf{Panel b:} the same as panel a but shows the image after subtracting static sources. Despite the presence of nearby stars, the asteroid is detected cleanly in the subtracted images. \textbf{Panel c:} the same as panel a, but the second of four o-band images containing \pt taken on 2024 August 7 21:16:35 UTC. \textbf{Panel d:} the same as panel c but shows the image after subtracting static sources. The asteroid was detected with an apparent magnitude of o=17.13 in panel b and o=16.99 in panel c.  The large black areas in the subtracted images are regions of saturated pixels from bright stars.  The direction of the asteroid and cardinal directions are indicated in each figure.}
\end{figure}

On 2024 September 27 05:54 UTC, \pt was observed using the 8.1 m Gemini North telescope with the Gemini Multi-Object Spectrograph (GMOS) instrument \citep[][]{Hook2004} in imaging mode (Program ID GN-2024B-FT-106, PI B. Bolin). GMOS was used with the Hamamatsu array with an effective pixel scale of 0.0807\arcsec/pixel in $2\times2$ binning mode. The images were taken in Sloan Digital Sky Survey (SDSS)-equivalent g, r, and i filters \citep[effective wavelengths 475 nm, 630 nm, and 780 nm, ][]{Fukugita1996}. For coverage of the spectrum of \pt past 800 nm, we use the GMOS-N Z-band filter, which is effectively equivalent to the United Kingdom Infrared Telescope Wide-Field Camera (WFCAM) Z filter \citep[][]{Hewett2006}. The WFCAM Z filter \citep[effective wavelength 880 nm, ][]{Casali2007} was designed to be equivalent to the SDSS z filter (effective wavelength 890 nm) but avoids the atmospheric absorption band near 950 nm \citep[][]{Hodgkin2009}.

The observations of \pt with GMOS occurred when it was near RA = 16 59 59.22, dec = +67 01 25.5, 8.74 lunar distances from the Earth (0.0225 au), a heliocentric distance of 1.000 au, and had a phase angle of 84.4$^{\circ}$. The seeing was $\sim$0.5-0.6\arcsec\ during the \pt observations as measured in g band images, and airmass ranged from 1.73 -1.91 during the $\sim$40 minute observing sequence. During the sequence, we acquired 12 x 75 s g band exposures, 5 x 50 s r band exposures, 3 x 60 s i band exposures, and 3 x 90 s  Z band exposures. Observations in the g, r, i, and Z filters were interspersed throughout the observation sequence to mitigate the effects of brightness variations in \ptns's lightcurve on the color measurements. The telescope was tracked at the $\sim$93\arcsec/h rate of \pt during the exposures. Bias and flat frames were taken and used to detrend the images using the DRAGONS image pipeline \citep[][]{Labrie2023}. Images of \pt when it intersected a background star were discarded. A mosaic of the separate g, r, i, and Z median-combined composite stacks of the \pt images are shown in Fig~2. 

\begin{figure}
\centering
\includegraphics[scale=.27]{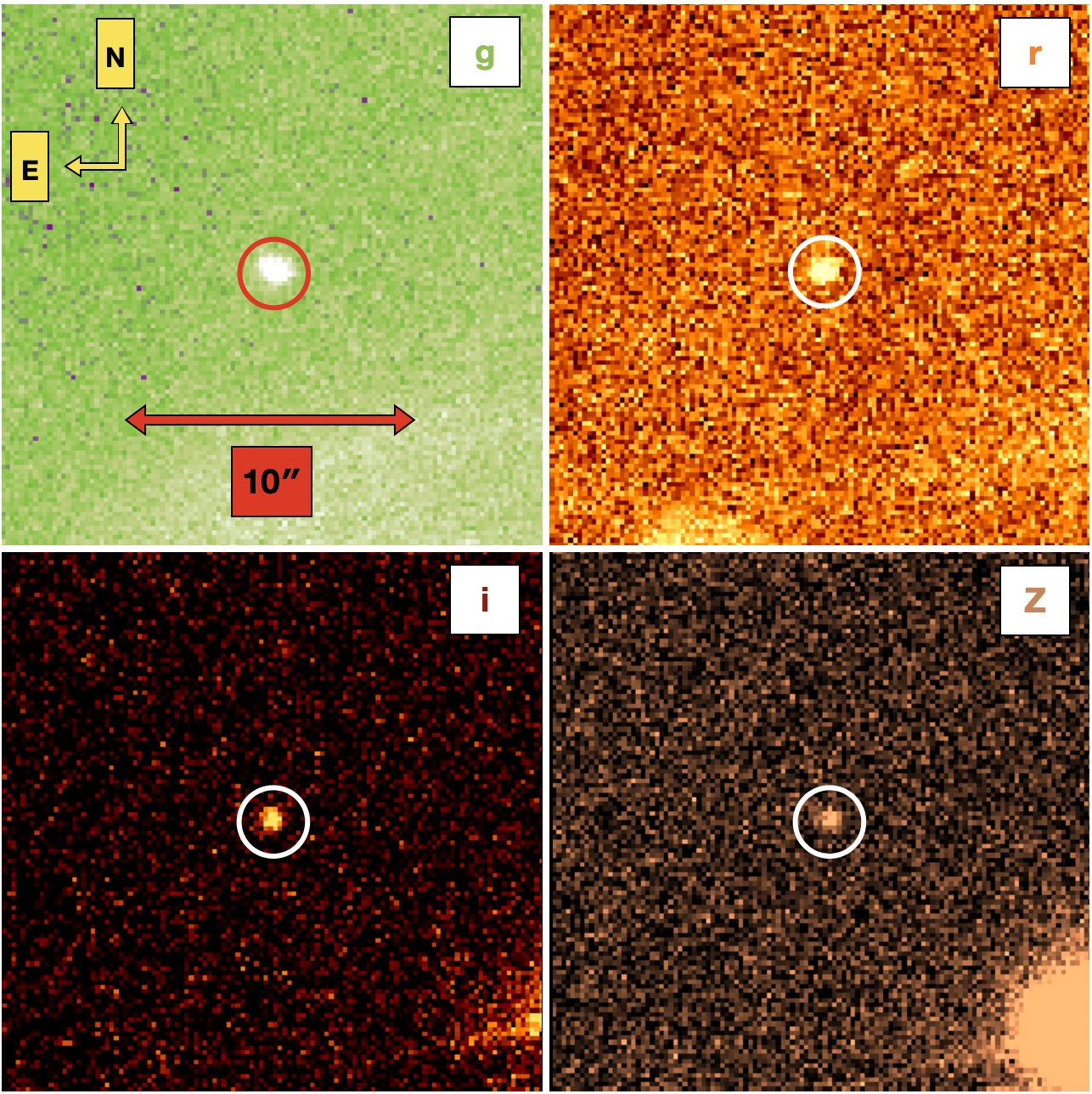}
\caption{\textbf{Top left panel:} a median combination stack of 11 x 75 s g filter images of \ptns. An arrow indicating the width of 10\arcsec~is shown for scale, and the cardinal directions are indicated. \textbf{Top right panel:} a median combination stack of 5 x 50 s r filter images of \ptns. \textbf{Bottom left panel:} a median combination stack of 2 x 60 s i filter images of \ptns. \textbf{Bottom right panel:} a median combination stack of 2 x 60 s Z filter images of \ptns. The image scale and cardinal in the r, i, and Z stacks are the same as in the g image stack.}
\end{figure}

\section{Results}
\subsection{Astrometry and orbital determination}
The astrometry from the Gemini North observations of \pt was measured with the Astrometrica software \citep[][]{Raab2012a} combined with reference stars from the \textit{Gaia} data release 2 catalog \citep[][]{Gaia2016,Gaia2018}.  We conservatively estimate an astrometric uncertainty of 1.0\arcsec~in both the right ascension and declination directions in line with the typical astrometric precision for most ground-based measurements \citep[e.g.,][]{Farnocchia2022}. Additionally, the timing of the GMOS instrument is accurate to within a few tenths of a millisecond, inducing negligible effect on the astrometry \footnote{\url{https://www.gemini.edu/observing/phase-iii-retrieving-reducing-data/reducing-data/timing-information}}.

The astrometry measured from the GMOS images was submitted to the MPC \citep[][]{Williams2024PT5}. The GMOS astrometry of \pt was combined with 200 additional \pt observations taken independently by follow-up observers between 2024 August 7 UTC and 2024 October 24 UTC from the MPC archive\footnote{\url{https://minorplanetcenter.net/db_search/show_object?utf8=\%E2\%9C\%93&object_id=2024+PT5}, accessed on 2024 November 7.}. A complete table of the astrometry of \pt is provided in Table~A1 of the appendix. We adopt conservative estimates for the astrometric uncertainties of $\sim$1.0\arcsec~in both right ascension and declination for the other observatories' measured positions except for observations made at the Great Shefford Observatory in West Berkshire, England (MPC observatory code J95), where we used an estimated uncertainty of 0.4\arcsec\ \citep[][]{Veres2017}.

Using the \texttt{Find$\_$Orb} orbital determination software by Bill Gray\footnote{\url{https://www.projectpluto.com/find_orb.htm}}, we fit an orbit to the observations in Table~A1, using the 8 planets and the Moon as perturbers, and including the effects of solar radiation pressure \citep[e.g.,][]{Fedorets2020CD3,Jewitt2024NonGrav}. Our orbital fit results in estimates of the heliocentric and geocentric semi-major axis, $a$, eccentricity, $e$, inclination, $i$, ascending node, $\Omega$, argument of perihelion, $\omega$, mean anomaly, $M$, and the area-to-mass ratio (AMR) as a measure of the effect of solar radiation pressure on the orbit of \ptns.

The heliocentric, geocentric orbital parameters and area-to-mass ratio  ($a$, $e$, $i$, $\Omega$, $\omega$, $M$, AMR) parameter estimates of the least square orbital fit using the observations of \pt is shown in Table~2. The mean observed-minus-computed residual from the least square fit using the 7-parameter fit is 0.33\arcsec. A complete list of the observed-minus-computed residuals in the along-track, X$_{res.}$, and cross-track, Y$_{res.}$, are shown in Table~A1. The geocentric eccentricity $e_g$ of \pt was $\sim$1.05 on 2024 September 27 UTC,  the date of the Gemini N observations, indicating that \pt was on a hyperbolic trajectory with respect to Earth. Its AMR of 7.02$\pm$2.05$\times$10$^{-5}$ m$^2$/kg is comparable to other small asteroids with measured AMRs \citep[e.g.,][]{Micheli2013,Jedicke2018mm}. 

The orbital trajectory of \pt between 2024 June and 2025 April is shown in Fig.~3 based on values from JPL Horizons\footnote{\url{https://ssd.jpl.nasa.gov/horizons/app.html\#/}}.  The $e_g$ of \pt decreased below 1.0 at around 2024 September 29 23:40 UTC, as indicated by the red portion of the trajectory of \pt in the top and bottom panels of Fig.~3. The capture of \pt by the Earth-Moon system ended at around 2024 November 25 18:20, making it the most short-lived minimoon known, having been captured by the Earth-moon system for only $\sim$60 days. Additionally, \pt did not make a complete revolution around the Earth-Moon barycenter, unlike previous minimoons such as 2006 RH$_{120}$ \citep[][]{Kwiatkowski2008,Granvik2012} and 2020 CD$_3$.

\begin{figure}
\centering
\includegraphics[scale=0.55]{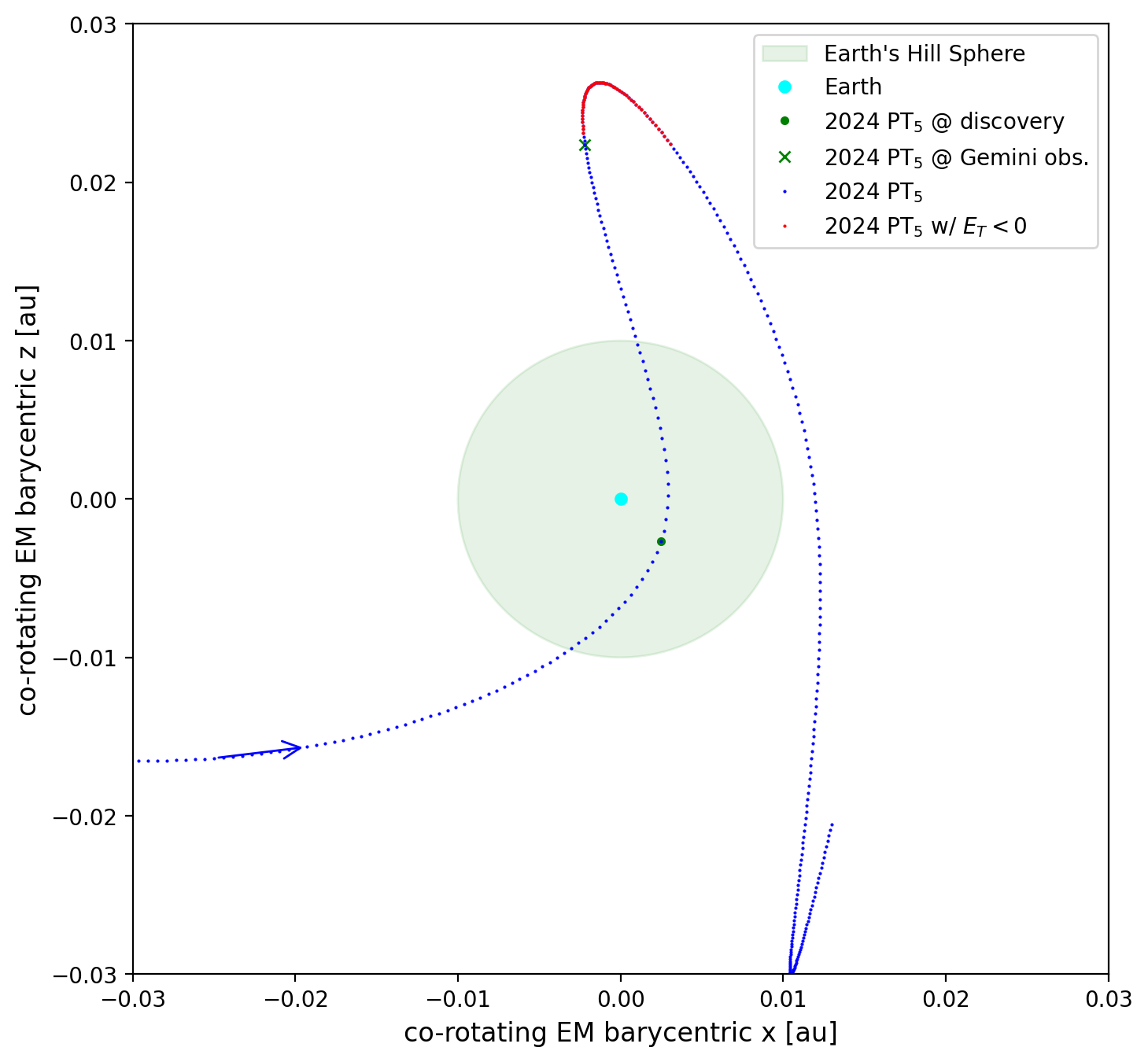}
\includegraphics[scale=0.55]{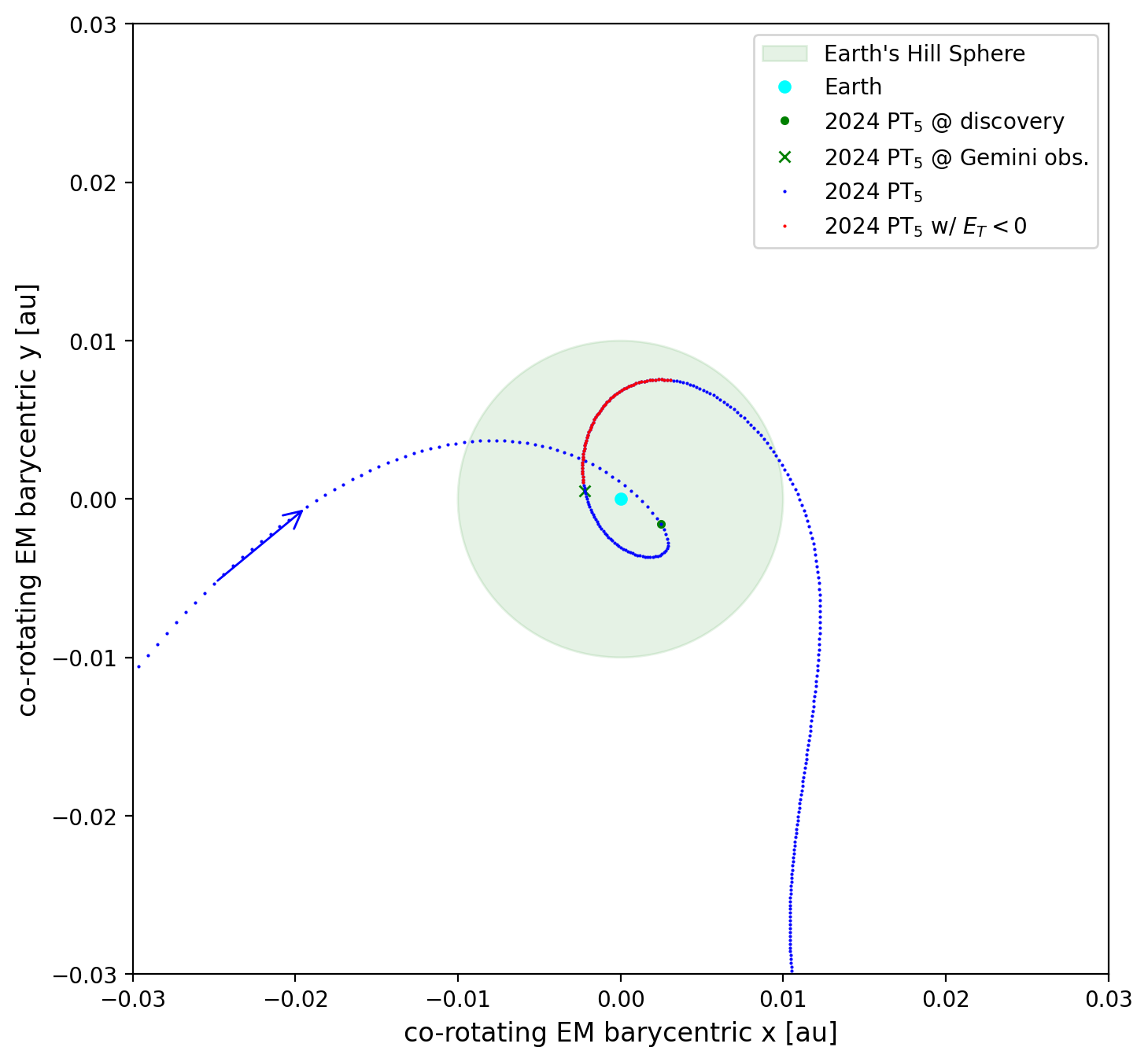}
\caption{\textbf{Top panel:} side view of the Earth co-rotating frame orbital trajectory of \pt as it enters and leaves the Earth-Moon system between 2024 June and 2025 April in Cartesian Earth-Moon barycentric x and z coordinates. The daily position of \pt is represented as blue points except the portion of its trajectory when it had $e_g$$<$1 plotted in red. The position of the Earth is plotted with a cyan circle. The circular green-shaded region indicates the Hill sphere of Earth. The position of \pt, when it was discovered, is marked by a green circle, and the position of \pt when Gemini observed it is marked with a green X. A blue arrow indicates the direction of motion of \pt along its orbital path. \textbf{Bottom panel:} the same as the top panel, except in Cartesian Earth-Moon barycentric x and y coordinates.}
\end{figure}

\subsection{Photometry and spectral classification}
\label{sec:photo}
The photometry of \pt was measured using a 0.81\arcsec~aperture and then subtracting from it the median contribution from the sky background within a  1.3-2.4\arcsec~annulus. The g, r, i, and Z photometry were calibrated using solar analog stars from the Pan-STARRS catalog \citet[][]{Chambers2016}. The Pan-STARRS catalog magnitudes of the solar analog stars were transformed to SDSS magnitudes using the conversions from \citet[][]{Tonry2012}. We obtained brightnesses of g = 23.52 $\pm$ 0.03, r = 22.95 $\pm$ 0.03, i = 22.66 $\pm$ 0.03, Z = 22.93 $\pm$ 0.05. The multi-band colors of \pt are g-r = 0.58$\pm$0.05, r-i = 0.29$\pm$0.04, and i-z = -0.27$\pm$0.06. The g-i color and spectra slope are 0.87$\pm$0.04 and 9.5$\pm1.1\%$/100 nm.

Using the definition of a$^{*}$ from  \citet[][]{Ivezic2001}, a$^{*}$ = (0.89 (g-r)) + (0.45 (r-i)) - 0.57, as an indicator of spectral slope, we find a$^{*}$ = 0.07$\pm$0.04 as plotted vs. i-z(Z) in Fig.~4. Overall, \pt has a$^{*}$ and  i-z similar to S-types which have a$^{*}$ = 0.12$\pm$0.03 and i-z = -0.08 $\pm$ 0.07 on average and some V-type asteroids which have a$^{*}$ = 0.15 $\pm$ 0.11 and i-z = -0.46 $\pm$ 0.04 on average as plotted in Fig.~4. \pt is redder compared with a smaller i-z color compared to C-type asteroids, which have on average an a$^{*}$ = -0.09$\pm$0.02 and i-z = 0.02$\pm$0.03 and the Sun, which has an a$^{*}$ = -0.13 and i-z = 0.03.

\begin{figure}
\centering
\includegraphics[scale=.405]{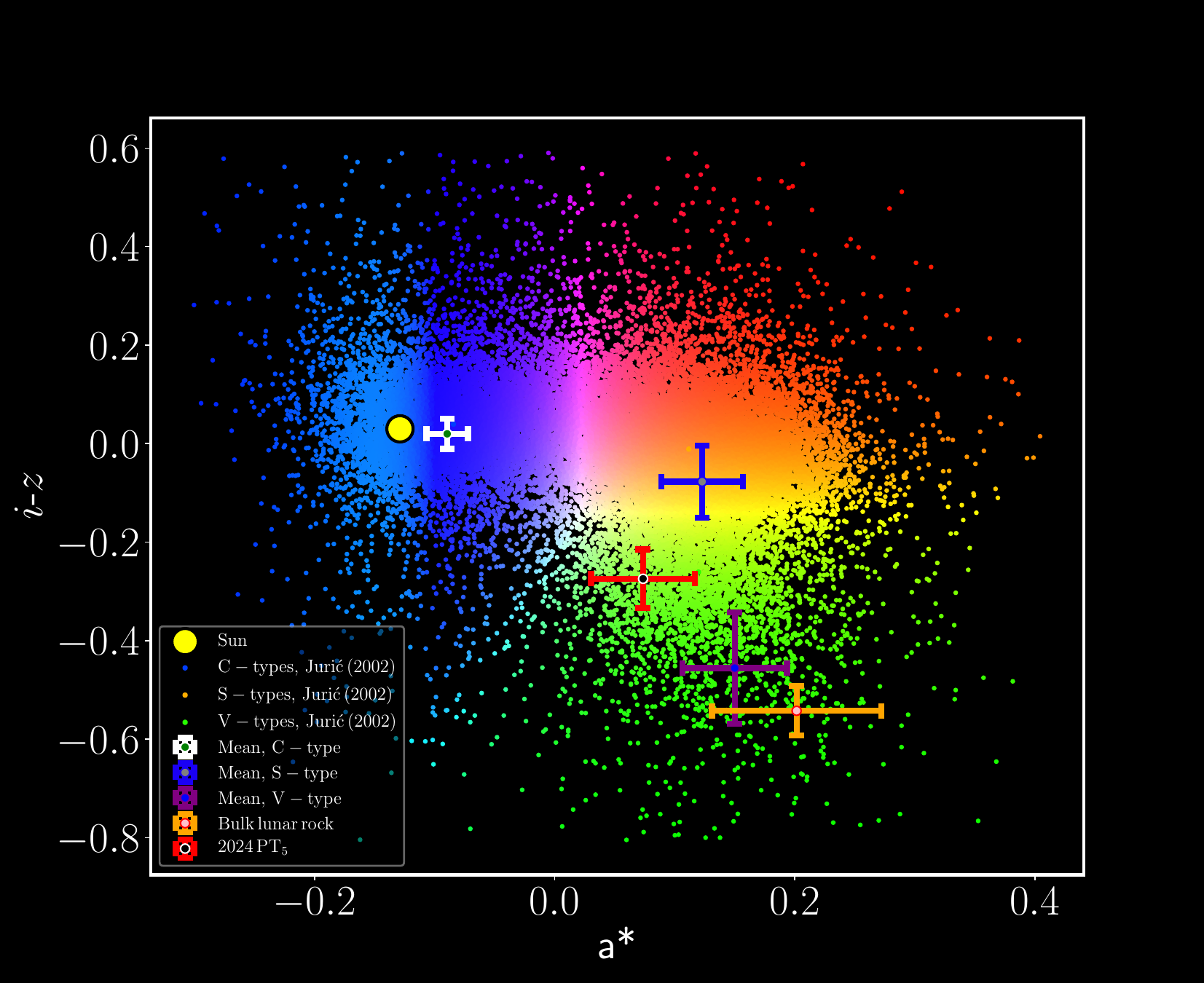}
\caption{a$^{*}$ vs. i-z(Z) colors of \pt plotted with a$^{*}$ vs. i-z colors of C, S and V type asteroids from \citep{Ivezic2001,Juric2002}, active comets \citep[][]{Solontoi2012}, Kuiper Belt Objects \citep[][]{Ofek2012}, and bulk lunar rock samples \citep[][]{Isaacson2011}. The colorization scheme of data points as a function of a$^{*}$ and i-z is adapted from \citet[][]{Ivezic2002} where blue symbol colors correspond to C-type asteroids, red symbol colors correspond to S-type asteroids and green symbol colors correspond to V-type asteroids. We note that in this case the measured Z magnitude of \pt is plotted as a substitute for its z magnitude. $i$-$z$. The a$^{*}$ and $i$-$z$ range of average S, V, and C-type asteroids are shown computed from the average spectra from \citet[][]{DeMeo2009}.}
\end{figure}

We compute the spectral reflectance of \pt by dividing its flux per g, r, i, and Z filter obtained by the flux of a solar analog in a corresponding filter, normalizing it to a wavelength of 550 nm. The normalized reflectivity spectrum of \pt shown in Fig.~5  is similar to the spectra of S- and V-type asteroids \citep[][]{Bus2002, DeMeo2009} with a red slope between 470 nm and 770 nm and a decrease in reflectance near 880 nm. The spectrum of \pt is also similar to some bulk basaltic lunar rock samples consisting of pyroxene minerals \citep[][]{Isaacson2011}. For a thorough comparison, we compute the $\chi^2$ statistic for the spectrum of \pt compared to the spectrum of 19 different asteroid types including S-complex asteroid types (S, Sq, Sv, Q),  C-complex types (B, C, Cg, Cgh), X-complex types (X, Xc, Xe, Xk, Xn), and other assorted types (A, D, K, L, O, R, V) from \citep[][]{Bus2002,DeMeo2009}. We find the closest match is to lunar rock with a reduced $\chi^2$ of 2.76 and Sv-type asteroids with a reduced $\chi^2$ of 3.75. By comparison, the comparison with V-type spectra results in a reduced $\chi^2$ of $\sim$12.84.

\begin{figure}
\centering
\includegraphics[scale=.38]{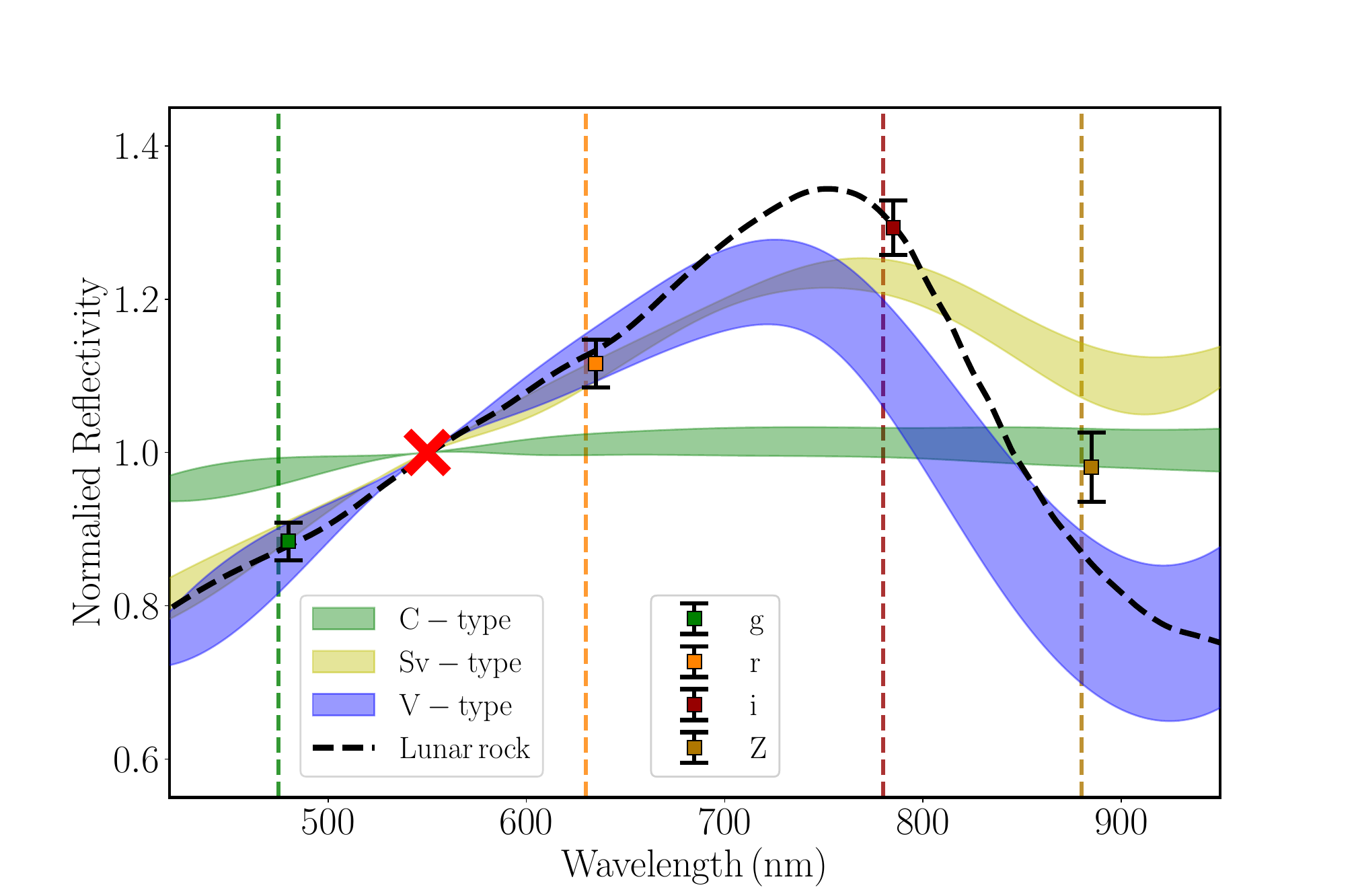}
\caption{Reflectance photometric spectrum of \pt consisting of g, r, i, and Z observations of \pt on 2024 September 27 UTC. The $\lambda_{\mathrm{eff}}$ locations of the g, r, i, and Z filters have been plotted as vertical dashed lines. The data points for the normalized reflectivity of \pt have been offset slightly from their location in the wavelength direction. The error bars on the spectrum data points correspond to 1$\sigma$ uncertainty. The spectrum has been normalized to unity at 550 nm, as indicated by the red cross. The spectral range of S, V and C-type asteroids from the Bus-DeMeo asteroid taxonomic catalog \citep[][]{DeMeo2009} are over-plotted with the V-type spectrum most closely resembling the spectra of \pt. The average spectrum of Apollo program coarse bulk lunar rock samples is plotted for reference \citep[][]{Isaacson2011}.}
\end{figure}

We use our g and r magnitude measurements of \pt to estimate an equivalent V-band brightness of 23.2$\pm$0.04 on 2024 September 7.  We calculated the H magnitude of \pt using our estimated $V$ magnitude and the phase function from equation from \citet[][]{Bowell1988}:
\begin{equation}
\label{eqn.brightness}
H = V - 5\, \mathrm{log_{10}}(r_h \Delta) +2.5\,\mathrm{log_{10}}\left[ (1 - G)\,\Phi_1(\alpha) + G\,\Phi_2(\alpha) \right ]
\end{equation}
where $r_h$ is the 1.000 au heliocentric distance, $\Delta$ is its geocentric distance of 0.0225 au and $\alpha$ is its phase angle of 84.4$^\circ$ of \pt on 2024 September 27 UTC. $G$ is the phase coefficient which we use the value of 0.2, the average value for S-type asteroids \citep[][]{Veres2015}. $\Phi_1(\alpha)$ and $\Phi_2(\alpha)$ are the basis functions normalized at $\alpha$ = 0$^\circ$ described in \citep[][]{Bowell1988}. We obtain $H$ = 28.64$\pm$0.04 but caution that the uncertainty on H is underestimated due to the lack of information about the phase function.

\subsection{Lightcurve, periodicity, and axial ratio}
\label{s.lightcurve}

We use the g filter data to search for periodic variations in the time series brightness of \ptns. The measured time series photometric values in the individual g images are presented in Table~2 and plotted in the top panel of Fig.~6. There are variations as large as $\sim$0.2-0.25 magnitudes in the brightness of \pt, significantly larger than the $\sim$0.01 magnitude variations seen in Skyprobe data for 2024 September 27 UTC\footnote{\url{https://www.cfht.hawaii.edu/cgi-bin/elixir/skyprobe.pl?plot&mcal_20240927.png}} and the $\sim$0.05 mag uncertainty of the individual g filter data points. 

\begin{figure}
\centering
\includegraphics[scale=0.45]{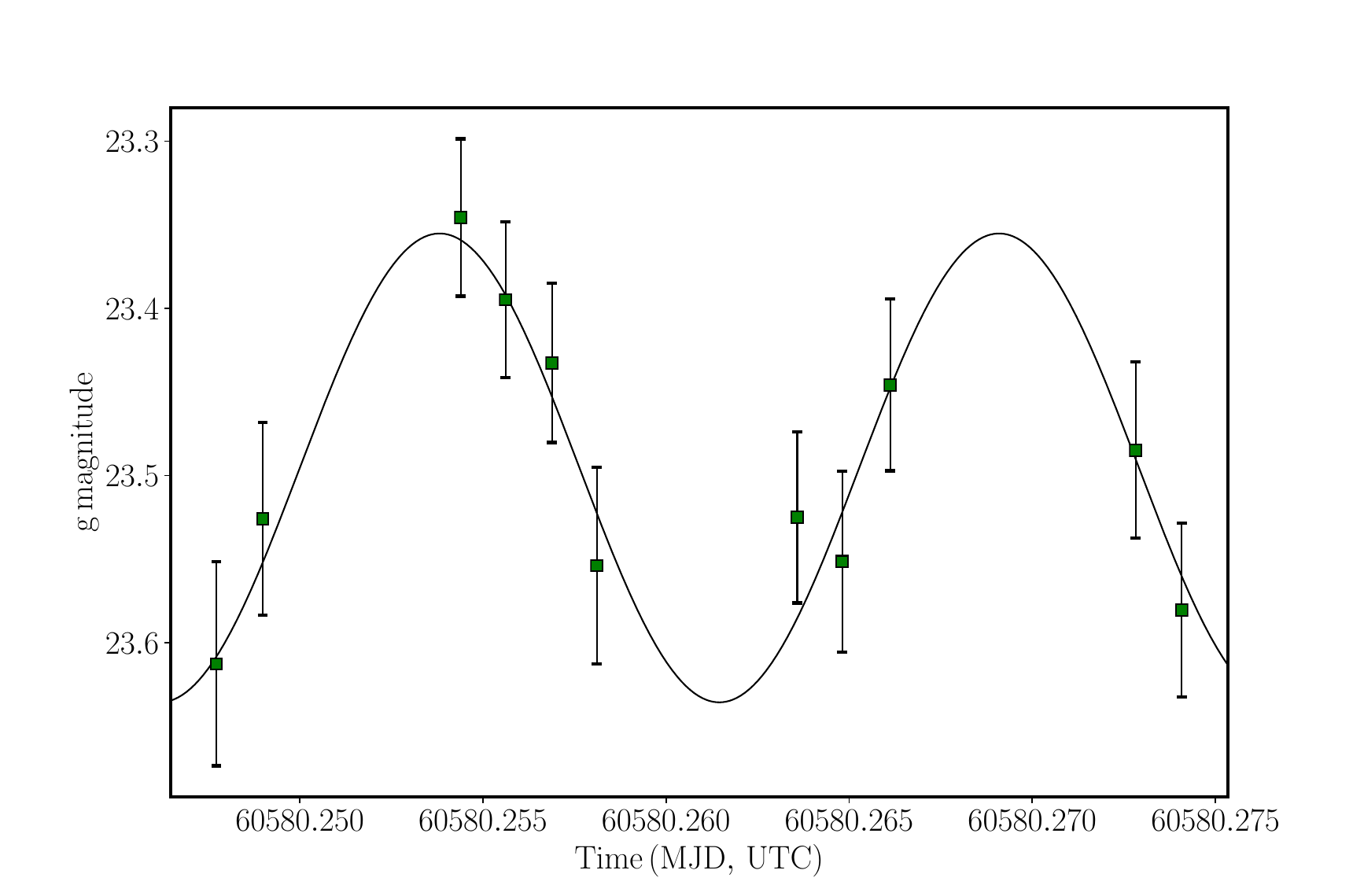}
\includegraphics[scale=0.45]{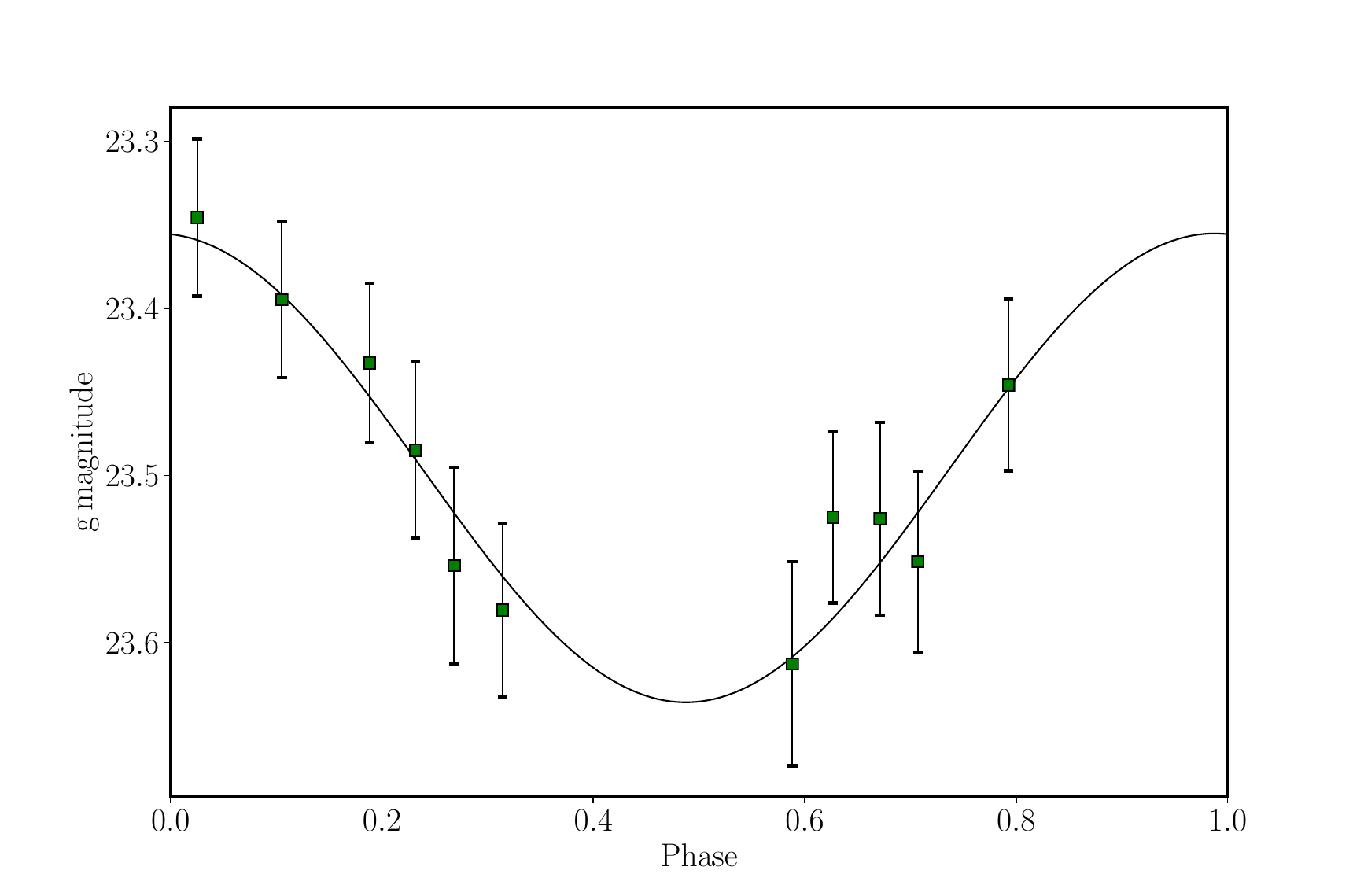}
\caption{\textbf{Top panel:} g filter lightcurve from 2024 September 27 GMOS observations of \ptns. The error bars on the data points are equal to their 1 $\sigma$ photometric uncertainties. A model lightcurve is plotted in black with a double-peaked period of $\sim$2600 s and amplitude of 0.28 magnitudes. \textbf{Bottom panel:} phased g filter lightcurve data of \pt using a lightcurve period of $\sim$1300 s and amplitude of 0.28 magnitude.}
\end{figure}

The Lomb-Scargle periodogram \citep[][]{Lomb1976} was applied to the time series g filter data, as seen in the top panel of Fig.~7. We take a similar approach as \citet[][]{Bolin2024Streak} to estimating the uncertainty on the determined period by removing $\sqrt{N}$ data points from the time series lightcurve and repeating our periodogram estimation of the lightcurve period 10,000 times, resulting in a central value of $\sim$1317 s and a 1~$\sigma$ uncertainty estimate of $\sim$227 s. We apply phase dispersion minimization analysis to our data \citep[][]{Stellingwerf1978} as an independant check of the Lomb-Scargle results, obtaining a result of $\sim$1150 s as seen in the bottom panel of Fig.~7, comparable with the lightcurve period estimate. These period estimates imply that \pt has a double-peaked rotation period of $\sim$2600$\pm$500 s.

\begin{figure}
\centering
\includegraphics[scale=0.45]{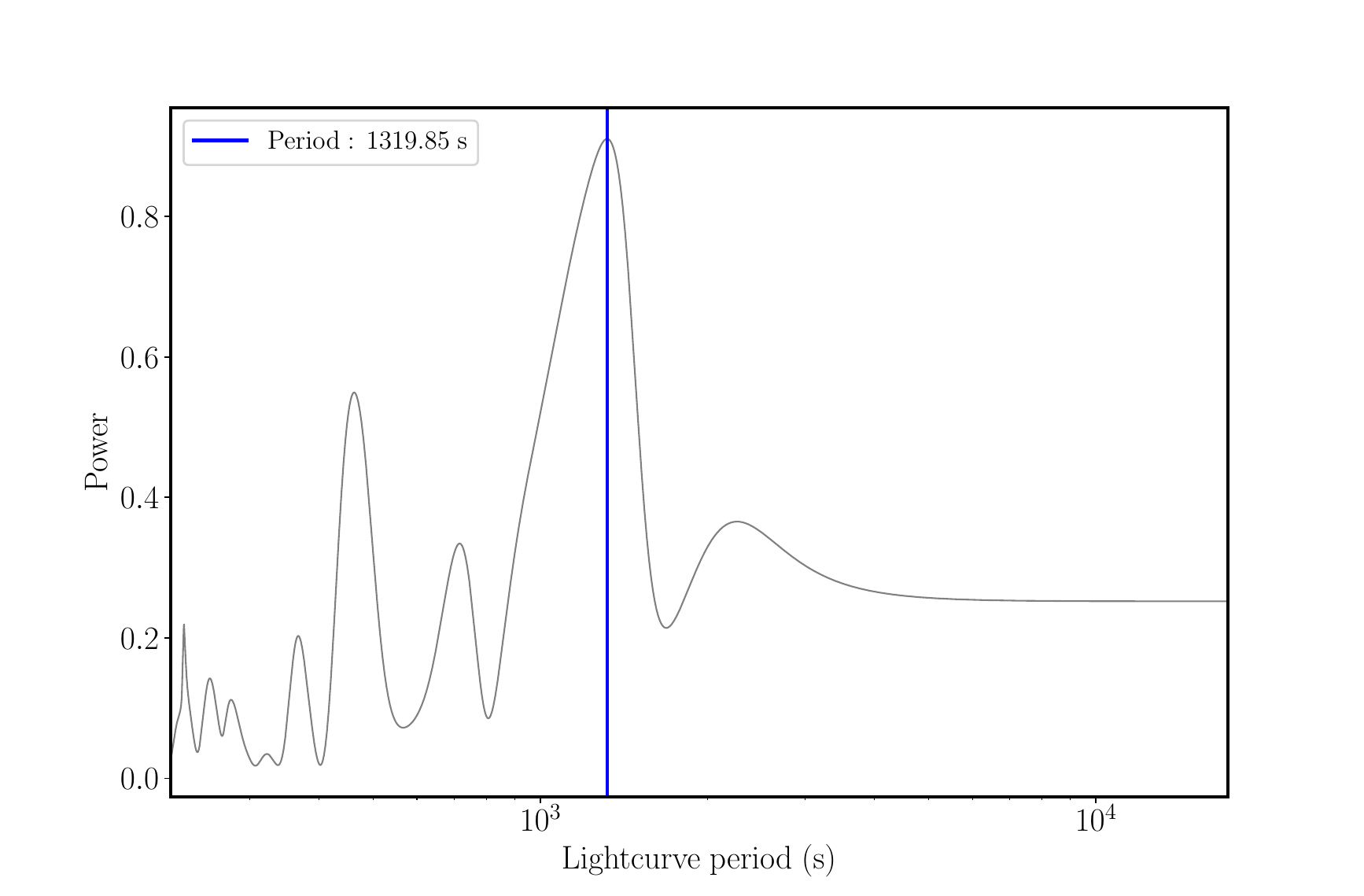}
\includegraphics[scale=0.45]{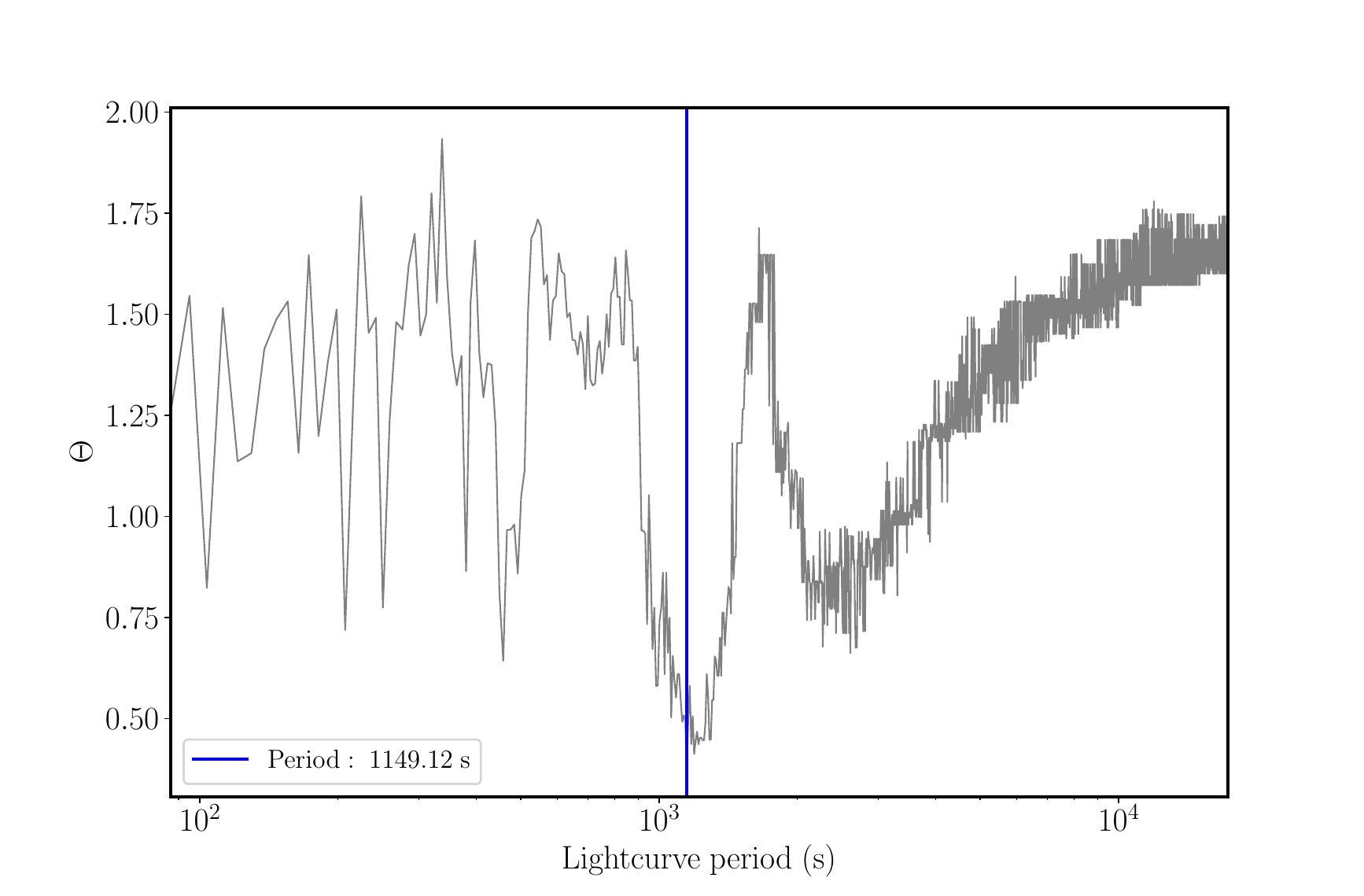}
\caption{Top panel: Lomb-Scargle periodogram of lightcurve period vs. spectral power \citep[][]{Lomb1976} for the g filter data from the 2024 September 27 UTC Gemini N/GMOS observations. A peak in the power is located at single-peaked lightcurve period of 1320 s. Bottom panel: Phase dispersion minimization analysis of lightcurve rotation period vs. $\Theta$ metric \citep[][]{Stellingwerf1978}. The $\Theta$ metric is minimized at single-peaked rotation periods of 1150 s similar to the 1320$\pm$227 s period found with the Lomb-Scargle Periodogram.}
\label{fFiig:periodogram}
\end{figure}

Using the observed amplitude of the g filter lightcurve of 0.28$\sim$0.07, we estimate the rough shape of \pt assuming it is a prolate triaxial ellipsoid with dimensions a:b:c, where b $\geq$ a $\simeq$ c in rough agreement with the shapes of other asteroids inferred from lightcurve inversion \citep[e.g.,][]{Hanus2016a, Hanus2018}.  The ratio between $b/a$ is described by $b/a \; = \; 10^{0.4 A}$ where $A$ is the peak-to-trough lightcurve amplitude \citep[][]{Binzel1989} and assuming a $\simeq$ c, resulting in a $b/a\sim$1.3. However, the axial ratio may be exaggerated due to the effect of observing at a large phase angle of $\sim$84$^{\circ}$ on 2024 September 27 UTC \citep[][]{Zappala1990a} as has been suggested for lightcurve observations of other small solar system objects \citep[e.g.,][]{Bolin2018}.

\section{Discussion and Conclusions}

The spectrum of \pt is similar to asteroids from the S-complex, as well as some samples of lunar rock, which suggests that it may either originate from the inner Main Belt, where the majority of S-type asteroids are found \citep[][]{DeMeo2014a}, or as a piece of debris resulting from an impact on the moon ejected into an Earth-similar orbit \citep[][]{Gladman1995}. We can use the orbital information of \pt as an additional constraint on its likely origin and physical properties through comparison with dynamical models describing the escape of asteroids from the Main Belt into the NEO population \citep[][]{Granvik2017,Nesvorny2023NEOMOD}.

We use the visible albedos, $p_v$, measured for S-complex and lunar rock samples to estimate the albedo of \ptns. S-complex asteroids have an albedo of $\sim$0.21 \citep[][]{Thomas2011}, while lunar rock has an albedo of $\sim$0.14 \citep[][]{Matthews2008}. As an independent albedo estimate, we compare the orbit of \pt with the NEOMOD3 population model suggests its most likely source is the $\nu_6$ resonance located near the inner edge of the Main Belt with an 88.4$\%$ probability. The next most likely location, with a 11$\%$ chance of \pt coming from it, is the 3:1 mean motion resonance located in the Main Belt at 2.5 au.  Weighing the NEO albedo model \citep[][]{Morbidelli2020albedo} according to these source probabilities for \pt results in a predicted $p_v$ of $\sim$0.21. We assuming a conservative albedo estimate of $\sim$0.1 corresponding to the scatter in albedo measurements for small S-complex asteroids \citep[][]{Delbo2003, Binzel2004}.

We use the albedo estimate and H magnitude of \pt to determine the diameter, D, of \pt using the equation $\mathrm{D = \frac{1329}{\sqrt{p_v}}10^{-\frac{H}{5}}}$ from \citet[][]{Russell1916} arriving at D = 5.4$\pm$1.2 m. Combining this diameter calculation with the AMR determined from our orbital fit, we can estimate a bulk density for \pt of 3.9$\pm$2.1g/cm$^3$ comparable with achondritic basaltic meteorites \citep[][]{Macke2011} and small asteroids with AMR measurements \citep[][]{Mommert2014}. If we use the mean albedo of lunar rock samples of 0.14, we obtain an diameter of 6.0$\pm$1.7 m and a bulk density of 3.6$\pm$2.4g/cm$^3$. This density estimate is comparable to the density of some lunar rock samples \citep[][]{Kiefer2012}. Using these density estimates, the mass of \pt is $\sim$10$^5$ kg. With the comparison of its physical properties determined from our observations to the physical properties of asteroids and lunar samples, we conclude that \pt the physical properties of \pt deduced from our observations is compatible within an inner Main Belt or lunar ejecta origin.

\section*{acknowledgments}
This study is based on observations obtained at the international Gemini Observatory, a program of NSF's NOIRLab, which is managed by the Association of Universities for Research in Astronomy (AURA) under a cooperative agreement with the National Science Foundation on behalf of the Gemini Observatory partnership.

Gemini Observatory is located on Maunakea, land of the K$\mathrm{\bar{a}}$naka Maoli people, and a mountain of considerable cultural, natural, and ecological significance to the indigenous Hawaiian people. The authors wish to acknowledge the importance and reverence of Maunakea and express gratitude for the opportunity to conduct observations from the mountain.

\facility{Gemini North} 

\bibliographystyle{aasjournal}
\bibliography{/Users/bolin/Dropbox/Projects/NEOZTF_NEOs/neobib}

\begin{table}
\centering
\caption{Orbital elements of \pt  based on observations collected between 2024 August 7 UTC and 2020 October 24 UTC. The orbital elements are shown for the Julian date (JD) using the software \texttt{Find$\_$Orb} by Bill Gray. The 1~$\sigma$ uncertainties are given in parentheses.}
\label{t:orbit}
\begin{tabular}{ll}
\hline
Heliocentric Elements&
\\ \hline
Epoch (JD) & 2,460,607.5\\
\hline
Time of perihelion, $T_p$ (JD) & 2,460,638.2598770$\pm$(5.58x10$^{-5}$)\\
Semi-major axis, $a$ (au) & 1.0121184340$\pm$(2.03x10$^{-8}$)\\
Eccentricity, $e$ & 0.021315140$\pm$(3.89x10$^{-8}$)\\
Perihelion, $q$ (au) & 0.9905449810$\pm$(3.01x10$^{-8}$)\\
Aphelion, $Q$ (au) & 1.0336918860$\pm$(5.53x10$^{-8}$)\\
Inclination, $i$ ($^{\circ}$) & 1.5167380$\pm$(4.7x10$^{-6}$)\\
Ascending node, $\Omega$ ($^{\circ}$) & 304.494733$\pm$(9.0x10$^{-5}$)\\
Argument of perihelion, $\omega$ ($^{\circ}$) & 117.54969$\pm$(1.0x10$^{-4}$)\\
Mean Anomaly, $M$ ($^{\circ}$) & 330.225693$\pm$(5.4x10$^{-5}$)\\
\hline
Geocentric Elements&\\
\hline
Epoch (JD) & 2,460,607.5\\
\hline
Time of perihelion, $T_{p,g}$ (JD) & 2,460,429.570191$\pm$(2.87x10$^{-4}$)\\
Semi-major axis, $a_g$ (au) & 0.017623959$\pm$(1.09x10$^{-7}$)\\
Eccentricity, $e_g$ & 0.61405662$\pm$(2.04x10$^{-6}$)\\
Perihelion, $q_g$ (au) & 0.006801850$\pm$(7.1x10$^{-8}$)\\
Aphelion, $Q_g$ (au) & 0.028446068$\pm$(1.56x10$^{-7}$)\\
Inclination, $i_g$ ($^{\circ}$) & 105.640940$\pm$(4.0x10$^{-5}$)\\
Ascending node, $\Omega_g$ ($^{\circ}$) & 302.876944$\pm$(3.40x10$^{-4}$)\\
Argument of perihelion, $\omega_g$ ($^{\circ}$) & 277.668305$\pm$(3.9x10$^{-4}$)\\
Mean Anomaly, $M_g$ ($^{\circ}$) & 129.9004$\pm$(1.0x10$^{-3}$)\\
\hline
Area-to-Mass ratio, AMR (m$^2$/kg)& 7.02x10$^{-5}$$\pm$(2.05x10$^{-5}$)\\
Absolute Magnitude, $H$ & 28.64$\pm$(0.04)\\
\hline
\end{tabular}
\end{table}

\clearpage
\newpage
\begin{longtable}{|c|c|c|c|}
\caption{Summary of \pt photometry taken on 2024 September 27 UTC.\label{t.photometry1}}\\
\hline
Date$^1$ & Filter$^2$ & Exp$^3$&$m^4$ \\
(MJD UTC)&&(s)&\\
\hline
\endfirsthead
\multicolumn{4}{c}%
{\tablename\ \thetable\ -- \textit{Continued from previous page}} \\
\hline
Date$^1$ & Filter$^2$ & Exp$^3$&$m^4$ \\
UTC&&(s)&\\
\hline
\endhead
\hline \multicolumn{4}{r}{\textit{Continued on next page}} \\
\endfoot
\hline
\endlastfoot
60580.2477197 & g & 75 s & 23.61 $\pm$ 0.06\\
60580.2489847 & g & 75 s & 23.53 $\pm$ 0.06\\
60580.2543947 & g & 75 s & 23.35 $\pm$ 0.05\\
60580.2556186 & g & 75 s & 23.39 $\pm$ 0.05\\
60580.2568848 & g & 75 s & 23.43 $\pm$ 0.05\\
60580.2581092 & g & 75 s & 23.55 $\pm$ 0.06\\
60580.2635833 & g & 75 s & 23.53 $\pm$ 0.05\\
60580.2648080 & g & 75 s & 23.55 $\pm$ 0.05\\
60580.2661196 & g & 75 s & 23.45 $\pm$ 0.05\\
60580.2728218 & g & 75 s & 23.48 $\pm$ 0.05\\
60580.2740833 & g & 75 s & 23.58 $\pm$ 0.05\\
\hline
\caption{Columns: (1) observation date correct for light travel time; (2) Gemini N/GMOS Filter; (3) Exposure time (4) per filter apparent magnitude with 1 $\sigma$ uncertainties}
\label{t:photo}
\end{longtable}

\clearpage
\newpage
\renewcommand{\thefigure}{A\arabic{figure}}
\setcounter{figure}{0}
\renewcommand{\thetable}{A\arabic{table}}
\renewcommand{\theequation}{A\arabic{equation}}
\renewcommand{\thesection}{A\arabic{section}}
\setcounter{section}{0}
\cleardoublepage
\setcounter{page}{1}
\renewcommand\thepage{A\arabic{page}}
\appendix
\renewcommand{\thefigure}{A\arabic{figure}}
\setcounter{figure}{0}
\renewcommand{\thetable}{A\arabic{table}}
\renewcommand{\theequation}{A\arabic{equation}}
\renewcommand{\thesection}{A}
\setcounter{section}{0}
\setcounter{table}{0}
\begin{longtable}{|c|c|c|c|c|c|c|c|}
\caption{Summary of astrometry from observations taken by Gemini North/GMOS and other observatories between 2024 August 7 UTC and 2024 October 24 UTC.\label{t.astrometry}}\\
\hline
Date$^1$ & R.A$^2$ & Dec.$^3$&$\sigma_{\mathrm{R.A.}}$$^4$&$\sigma_{\mathrm{Dec.}}$$^5$&$X_{res.}$$^6$&$Y_{res.}$$^7$&Obs. code$^8$\\
(UTC)&&&(\arcsec)&(\arcsec)&(\arcsec)&(\arcsec)&\\
\hline
\endfirsthead
\multicolumn{4}{c}%
{\tablename\ \thetable\ -- \textit{Continued from previous page}} \\
\hline
Date$^1$ & R.A$^2$ & Dec.$^3$&$\sigma_{\mathrm{R.A.}}$$^4$&$\sigma_{\mathrm{Dec.}}$$^5$&$X_{res.}$$^6$&$Y_{res.}$$^7$&Obs. code$^8$\\
(UTC)&&&(\arcsec)&(\arcsec)&(\arcsec)&(\arcsec)&\\
\hline
\endhead
\hline \multicolumn{4}{r}{\textit{Continued on next page}} \\
\endfoot
\hline
\endlastfoot
2024	8	7.883302	&	18	57	04.97	&	-55	56	00.24	&	1	&	1	&	0.42	&	-0.3	&	M22\\
2024	8	7.886527	&	18	56	56.059	&	-55	53	40.85	&	1	&	1	&	0.3	&	-0.04	&	M22\\
2024	8	7.892126	&	18	56	40.711	&	-55	49	37.52	&	1	&	1	&	0.37	&	0.05	&	M22\\
2024	8	7.903165	&	18	56	10.685	&	-55	41	31.85	&	1	&	1	&	-0.64	&	-0.03	&	M22\\
2024	8	12.060358	&	18	14	04.848	&	-09	25	00.44	&	1	&	1	&	0.13	&	0.29	&	W68\\
2024	8	12.065029	&	18	13	59.82	&	-09	22	24.35	&	1	&	1	&	0.27	&	-0.4	&	W68\\
2024	8	12.070721	&	18	13	53.688	&	-09	19	13.58	&	1	&	1	&	0.28	&	-0.88	&	W68\\
2024	8	12.105146	&	18	13	17.028	&	-08	59	51.4	&	1	&	1	&	-0.23	&	0.13	&	W68\\
2024	8	12.319011	&	18	12	25.265	&	-07	31	12.58	&	1	&	1	&	0.01	&	-0.43	&	T08\\
2024	8	12.321781	&	18	12	22.212	&	-07	29	42.04	&	1	&	1	&	0.12	&	-0.56	&	T08\\
2024	8	12.329519	&	18	12	13.702	&	-07	25	27.95	&	1	&	1	&	0.11	&	0.04	&	T08\\
2024	8	12.341403	&	18	12	00.773	&	-07	18	58.54	&	1	&	1	&	0.03	&	-0.36	&	T08\\
2024	8	12.890515	&	18	09	43.9	&	-02	45	25.0	&	1	&	1	&	0.13	&	0.1	&	I93\\
2024	8	12.892993	&	18	09	41.83	&	-02	44	10.7	&	1	&	1	&	0.42	&	0.07	&	I93\\
2024	8	12.896212	&	18	09	39.07	&	-02	42	34.8	&	1	&	1	&	-0.38	&	-0.56	&	I93\\
2024	8	12.90611	&	18	08	55.59	&	-02	37	52.5	&	1	&	1	&	-0.19	&	0.48	&	M45\\
2024	8	12.90899	&	18	08	53.62	&	-02	36	25.8	&	1	&	1	&	-0.03	&	0.62	&	M45\\
2024	8	12.91187	&	18	08	51.68	&	-02	34	59.5	&	1	&	1	&	0.26	&	0.39	&	M45\\
2024	8	12.912952	&	18	09	30.362	&	-02	36	41.8	&	0.4	&	0.4	&	0.25	&	0.11	&	J95\\
2024	8	12.914243	&	18	09	29.386	&	-02	36	03.24	&	0.4	&	0.4	&	0.21	&	0.07	&	J95\\
2024	8	12.916158	&	18	09	27.936	&	-02	35	06.0	&	0.4	&	0.4	&	0.07	&	0.05	&	J95\\
2024	8	12.916981	&	18	09	27.326	&	-02	34	41.34	&	0.4	&	0.4	&	0.19	&	0.11	&	J95\\
2024	8	12.92071	&	18	08	49.26	&	-02	30	57.2	&	1	&	1	&	-0.11	&	0.52	&	L04\\
2024	8	12.92166	&	18	09	03.22	&	-02	29	37.2	&	1	&	1	&	-0.69	&	-0.06	&	126\\
2024	8	12.92453	&	18	08	46.74	&	-02	29	02.7	&	1	&	1	&	-0.28	&	0.5	&	L04\\
2024	8	12.925395	&	18	09	19.685	&	-02	30	30.96	&	1	&	1	&	0.01	&	0.24	&	Z80\\
2024	8	12.92612	&	18	08	45.73	&	-02	28	14.9	&	1	&	1	&	0.06	&	0.65	&	L04\\
2024	8	12.928026	&	18	09	00.72	&	-02	26	59.4	&	1	&	1	&	0.29	&	0.06	&	203\\
2024	8	12.92976	&	18	08	43.4	&	-02	26	26.1	&	1	&	1	&	0.18	&	0.42	&	L04\\
2024	8	12.930038	&	18	09	16.248	&	-02	28	13.04	&	1	&	1	&	-0.52	&	-0.48	&	Z80\\
2024	8	12.933006	&	18	08	57.61	&	-02	24	38.1	&	1	&	1	&	0.29	&	0.17	&	204\\
2024	8	12.934692	&	18	09	12.878	&	-02	25	53.69	&	1	&	1	&	-0.38	&	-0.04	&	Z80\\
2024	8	12.93511	&	18	08	53.09	&	-02	22	55.3	&	1	&	1	&	-0.16	&	0.4	&	126\\
2024	8	12.93542	&	18	08	55.23	&	-02	23	17.9	&	1	&	1	&	0.07	&	-0.11	&	203\\
2024	8	12.93627	&	18	08	55.21	&	-02	23	00.8	&	1	&	1	&	0.32	&	0.24	&	204\\
2024	8	12.939514	&	18	08	52.88	&	-02	21	23.5	&	1	&	1	&	0.22	&	-0.23	&	204\\
2024	8	12.942278	&	18	08	50.3	&	-02	19	53.6	&	1	&	1	&	0.29	&	0.08	&	203\\
2024	8	12.94275	&	18	08	50.56	&	-02	19	47.0	&	1	&	1	&	0.33	&	0.02	&	204\\
2024	8	12.945971	&	18	08	48.28	&	-02	18	10.9	&	1	&	1	&	0.05	&	-0.56	&	204\\
2024	8	13.01645	&	18	09	01.788	&	-01	10	9.24	&	1	&	1	&	-0.33	&	-0.11	&	Y05\\
2024	8	13.018163	&	18	09	00.034	&	-01	09	18.53	&	1	&	1	&	0.31	&	-0.24	&	Y05\\
2024	8	13.019828	&	18	08	58.4	&	-01	08	29.81	&	1	&	1	&	0.34	&	-0.07	&	Y05\\
2024	8	13.10167	&	18	08	34.805	&	+00	23	28.32	&	1	&	1	&	-0.28	&	0.71	&	W68\\
2024	8	13.104829	&	18	08	31.841	&	+00	21	56.3	&	1	&	1	&	0.11	&	-0.54	&	W68\\
2024	8	13.110256	&	18	08	26.88	&	+00	19	20.93	&	1	&	1	&	0.11	&	0.46	&	W68\\
2024	8	13.120185	&	18	08	17.878	&	+00	14	33.76	&	1	&	1	&	-0.14	&	0.88	&	W68\\
2024	8	13.179412	&	18	07	28.721	&	+00	13	41.99	&	1	&	1	&	-0.39	&	-0.15	&	W68\\
2024	8	13.179862	&	18	07	28.368	&	+00	13	53.76	&	1	&	1	&	0.01	&	-0.41	&	W68\\
2024	8	13.180313	&	18	07	28.058	&	+00	14	06.32	&	1	&	1	&	0.48	&	-0.2	&	W68\\
2024	8	130.4929	&	18	08	31.937	&	+00	09	03.1	&	1	&	1	&	0.73	&	-0.6	&	U52\\
2024	8	13.21124	&	18	08	30.293	&	+00	08	07.84	&	1	&	1	&	0.12	&	0.15	&	U52\\
2024	8	13.21319	&	18	08	28.594	&	+00	07	11.46	&	1	&	1	&	0.14	&	-0.95	&	U52\\
2024	8	13.239647	&	18	09	08.398	&	+00	15	40.75	&	1	&	1	&	-0.53	&	0.23	&	T08\\
2024	8	13.240107	&	18	09	07.91	&	+00	15	55.01	&	1	&	1	&	-0.49	&	-0.33	&	T08\\
2024	8	13.241031	&	18	09	07.02	&	+00	16	20.71	&	1	&	1	&	-0.06	&	-0.29	&	T08\\
2024	8	13.393218	&	18	06	40.975	&	+01	27	26.89	&	1	&	1	&	0.26	&	-0.47	&	T08\\
2024	8	13.393681	&	18	06	40.608	&	+01	27	39.46	&	1	&	1	&	0.11	&	0.24	&	T08\\
2024	8	13.394144	&	18	06	40.21	&	+01	27	52.92	&	1	&	1	&	0.02	&	-0.35	&	T08\\
2024	8	13.394605	&	18	06	39.818	&	+01	28	05.02	&	1	&	1	&	-0.07	&	-0.01	&	T08\\
2024	8	13.395066	&	18	06	39.427	&	+01	28	18.05	&	1	&	1	&	-0.22	&	-0.62	&	T08\\
2024	8	13.395503	&	18	06	40.692	&	+01	27	53.75	&	1	&	1	&	-0.61	&	0.37	&	T05\\
2024	8	13.395955	&	18	06	40.289	&	+01	28	07.18	&	1	&	1	&	0.38	&	-0.34	&	T05\\
2024	8	13.396407	&	18	06	39.979	&	+01	28	18.91	&	1	&	1	&	0.29	&	-0.99	&	T05\\
2024	8	13.396858	&	18	06	39.598	&	+01	28	30.68	&	1	&	1	&	0.36	&	-0.57	&	T05\\
2024	8	13.39731	&	18	06	39.228	&	+01	28	43.54	&	1	&	1	&	0.36	&	0.23	&	T05\\
2024	8	13.495137	&	18	07	16.651	&	+02	06	36.61	&	1	&	1	&	-0.16	&	0.02	&	Q06\\
2024	8	13.502696	&	18	07	10.0	&	+02	10	02.57	&	1	&	1	&	-0.28	&	0.16	&	Q06\\
2024	8	13.510252	&	18	07	03.417	&	+02	13	28.4	&	1	&	1	&	0.02	&	-0.01	&	Q06\\
2024	8	13.818888	&	18	06	10.31	&	+04	26	07.1	&	1	&	1	&	-0.18	&	-0.02	&	L01\\
2024	8	13.820447	&	18	06	09.1	&	+04	26	47.6	&	1	&	1	&	-0.11	&	-0.04	&	L01\\
2024	8	13.822003	&	18	06	07.91	&	+04	27	28.0	&	1	&	1	&	-0.26	&	0.42	&	L01\\
2024	8	13.8344	&	18	06	00.492	&	+04	30	42.59	&	1	&	1	&	-0.07	&	0.34	&	G19\\
2024	8	13.84207	&	18	05	55.073	&	+04	34	00.77	&	1	&	1	&	-0.29	&	-0.02	&	G19\\
2024	8	13.846848	&	18	05	53.06	&	+04	38	40.8	&	1	&	1	&	-0.14	&	0.25	&	104\\
2024	8	13.850708	&	18	05	50.09	&	+04	40	20.6	&	1	&	1	&	0.05	&	0.57	&	104\\
2024	8	13.85616	&	18	05	45.182	&	+04	40	04.01	&	1	&	1	&	0.03	&	-0.04	&	G19\\
2024	8	13.86987	&	18	05	35.688	&	+04	45	55.08	&	1	&	1	&	-0.54	&	0.32	&	G19\\
2024	8	13.89015	&	18	05	21.965	&	+04	54	32.83	&	1	&	1	&	-0.38	&	0.51	&	G19\\
2024	8	13.89554	&	18	05	18.427	&	+04	56	49.96	&	1	&	1	&	-0.28	&	0.09	&	G19\\
2024	8	13.99644	&	18	04	13.93	&	+05	41	28.8	&	1	&	1	&	-0.14	&	-0.13	&	K83\\
2024	8	13.9989	&	18	04	12.77	&	+05	42	28.8	&	1	&	1	&	-0.1	&	0.03	&	K83\\
2024	8	14.872873	&	18	02	20.237	&	+11	24	21.74	&	0.4	&	0.4	&	-0.21	&	-0.11	&	J95\\
2024	8	14.875193	&	18	02	18.746	&	+11	25	12.4	&	0.4	&	0.4	&	-0.14	&	-0.07	&	J95\\
2024	8	14.87675	&	18	01	52.49	&	+11	26	12.8	&	1	&	1	&	0.01	&	-0.1	&	G02\\
2024	8	14.876922	&	18	02	17.647	&	+11	25	50.27	&	0.4	&	0.4	&	0.1	&	-0.55	&	J95\\
2024	8	14.879154	&	18	02	16.234	&	+11	26	39.01	&	0.4	&	0.4	&	-0.15	&	-0.12	&	J95\\
2024	8	14.88029	&	18	01	50.38	&	+11	27	29.0	&	1	&	1	&	-0.02	&	0.1	&	G02\\
2024	8	14.937552	&	18	01	41.203	&	+11	51	18.89	&	1	&	1	&	-0.02	&	-0.15	&	Y89\\
2024	8	14.942425	&	18	01	37.93	&	+11	53	02.12	&	1	&	1	&	-0.19	&	0.45	&	Y89\\
2024	8	14.947234	&	18	01	34.727	&	+11	54	43.23	&	1	&	1	&	-0.06	&	0.57	&	Y89\\
2024	8	15.75102	&	18	00	18.23	&	+16	56	43.87	&	1	&	1	&	-0.24	&	0.32	&	M49\\
2024	8	15.75565	&	18	00	14.654	&	+16	58	16.18	&	1	&	1	&	-0.17	&	0.09	&	M49\\
2024	8	15.840247	&	17	59	17.45	&	+16	54	19.4	&	1	&	1	&	-0.26	&	-0.17	&	G34\\
2024	8	15.84308	&	17	59	19.699	&	+16	55	38.57	&	1	&	1	&	0.15	&	-0.3	&	K87\\
2024	8	15.847264	&	17	59	13.29	&	+16	56	29.4	&	1	&	1	&	-0.21	&	-0.36	&	G34\\
2024	8	15.849602	&	17	59	11.94	&	+16	57	12.6	&	1	&	1	&	-0.2	&	-0.32	&	G34\\
2024	8	15.851941	&	17	59	10.54	&	+16	57	55.8	&	1	&	1	&	0.24	&	0.03	&	G34\\
2024	8	15.85295	&	17	59	13.726	&	+16	58	42.13	&	1	&	1	&	-0.29	&	0.25	&	K87\\
2024	8	15.854281	&	17	59	09.20	&	+16	58	39.4	&	1	&	1	&	-0.03	&	-0.71	&	Z80\\
2024	8	15.86281	&	17	59	07.817	&	+17	01	45.08	&	1	&	1	&	0.03	&	0.57	&	G34\\
2024	8	16.908094	&	17	56	12.427	&	+22	05	37.28	&	1	&	1	&	0.42	&	-0.32	&	Z80\\
2024	8	16.919223	&	17	55	53.02	&	+22	07	56.8	&	1	&	1	&	-0.18	&	0.58	&	G34\\
2024	8	16.919319	&	17	56	06.643	&	+22	08	29.15	&	1	&	1	&	-0.25	&	0.13	&	Z80\\
2024	8	16.924262	&	17	55	50.71	&	+22	09	10.9	&	1	&	1	&	0.21	&	-0.1	&	G34\\
2024	8	16.934328	&	17	55	46.26	&	+22	11	37.5	&	1	&	1	&	0.12	&	-0.82	&	G34\\
2024	8	17.847128	&	17	54	06.09	&	+26	04	54.0	&	1	&	1	&	-0.25	&	0.18	&	203\\
2024	8	17.866413	&	17	53	55.23	&	+26	09	12.6	&	1	&	1	&	-0.06	&	0.73	&	203\\
2024	8	19.829362	&	17	49	49.51	&	+32	53	05.6	&	1	&	1	&	0.06	&	0.28	&	G34\\
2024	8	19.836813	&	17	49	45.94	&	+32	54	22.5	&	1	&	1	&	-0.61	&	-0.55	&	G34\\
2024	8	19.844249	&	17	49	42.41	&	+32	55	37.3	&	1	&	1	&	0.04	&	-0.19	&	G34\\
2024	8	19.851685	&	17	49	38.86	&	+32	56	50.8	&	1	&	1	&	-0.1	&	-0.06	&	G34\\
2024	8	23.848972	&	17	42	22.905	&	+42	36	29.87	&	1	&	1	&	-0.07	&	0.02	&	Z84\\
2024	8	23.855223	&	17	42	19.625	&	+42	37	09.02	&	1	&	1	&	-0.03	&	-0.01	&	Z84\\
2024	8	27.82145	&	17	35	51.09	&	+48	46	30.6	&	1	&	1	&	-0.14	&	0.11	&	033\\
2024	8	27.8226	&	17	35	50.65	&	+48	46	35.0	&	1	&	1	&	0.1	&	-0.01	&	033\\
2024	8	27.82374	&	17	35	50.2	&	+48	46	39.5	&	1	&	1	&	-0.1	&	-0.07	&	033\\
2024	8	28.837555	&	17	34	24.755	&	+50	04	53.76	&	1	&	1	&	-0.24	&	-0.01	&	Z84\\
2024	8	28.847621	&	17	34	20.044	&	+50	05	29.17	&	1	&	1	&	-0.26	&	0.26	&	Z84\\
2024	8	28.858796	&	17	34	14.908	&	+50	06	06.48	&	1	&	1	&	0.12	&	0.53	&	Z84\\
2024	9	10.85884	&	17	20	37.5	&	+60	13	29.6	&	1	&	1	&	0.26	&	0.01	&	Y65\\
2024	9	10.86154	&	17	20	36.47	&	+60	13	32.6	&	1	&	1	&	0.16	&	0.12	&	Y66\\
2024	9	10.88186	&	17	20	28.45	&	+60	13	48.9	&	1	&	1	&	0.04	&	0.54	&	Y65\\
2024	9	10.89336	&	17	20	24.45	&	+60	13	56.6	&	1	&	1	&	0.82	&	0.52	&	Y65\\
2024	9	10.89497	&	17	20	23.77	&	+60	13	57.4	&	1	&	1	&	0.28	&	0.49	&	Y64\\
2024	9	10.90274	&	17	20	21.01	&	+60	14	01.3	&	1	&	1	&	0.09	&	-0.35	&	Y64\\
2024	9	10.90364	&	17	20	20.68	&	+60	14	02.1	&	1	&	1	&	-0.04	&	0.25	&	Y66\\
2024	9	10.90487	&	17	20	20.36	&	+60	14	02.6	&	1	&	1	&	-0.23	&	0.17	&	Y65\\
2024	9	10.91735	&	17	20	16.11	&	+60	14	06.8	&	1	&	1	&	0.59	&	0.55	&	Y65\\
2024	9	10.92688	&	17	20	13.07	&	+60	14	08.0	&	1	&	1	&	0.35	&	0.18	&	Y64\\
2024	9	10.92983	&	17	20	12.27	&	+60	14	07.9	&	1	&	1	&	0.55	&	0.16	&	Y65\\
2024	9	10.93531	&	17	20	10.52	&	+60	14	09.6	&	1	&	1	&	0.28	&	0.24	&	Y64\\
2024	9	10.94153	&	17	20	08.71	&	+60	14	09.8	&	1	&	1	&	-0.04	&	-0.08	&	Y66\\
2024	9	10.94332	&	17	20	08.32	&	+60	14	10.2	&	1	&	1	&	-0.33	&	0.45	&	Y64\\
2024	9	10.95033	&	17	20	06.39	&	+60	14	09.6	&	1	&	1	&	0.04	&	-0.51	&	Y64\\
2024	9	10.95741	&	17	20	04.60	&	+60	14	08.9	&	1	&	1	&	0.02	&	0.41	&	Y64\\
2024	9	10.96505	&	17	20	02.72	&	+60	14	07.8	&	1	&	1	&	-0.12	&	0.09	&	Y64\\
2024	9	10.97207	&	17	20	01.09	&	+60	14	06.0	&	1	&	1	&	-0.13	&	0.11	&	Y64\\
2024	9	10.97949	&	17	19	59.49	&	+60	14	04.6	&	1	&	1	&	0.53	&	-0.45	&	Y64\\
2024	9	11.84545	&	17	19	52.21	&	+60	45	30.9	&	1	&	1	&	0.33	&	-0.09	&	Y65\\
2024	9	11.85686	&	17	19	47.59	&	+60	45	43.7	&	1	&	1	&	0.25	&	0.15	&	Y65\\
2024	9	11.86086	&	17	19	45.98	&	+60	45	47.1	&	1	&	1	&	0.29	&	-0.08	&	Y66\\
2024	9	11.87988	&	17	19	38.65	&	+60	46	01.8	&	1	&	1	&	-0.39	&	0.28	&	Y65\\
2024	9	11.89139	&	17	19	34.52	&	+60	46	07.8	&	1	&	1	&	-0.01	&	0.18	&	Y65\\
2024	9	11.89536	&	17	19	33.04	&	+60	46	12.0	&	1	&	1	&	-0.13	&	-0.03	&	Y64\\
2024	9	11.93403	&	17	19	20.61	&	+60	46	17.6	&	1	&	1	&	0.34	&	0.01	&	Y65\\
2024	9	11.9409	&	17	19	18.68	&	+60	46	18.7	&	1	&	1	&	0.4	&	-0.39	&	Y66\\
2024	9	11.97274	&	17	19	10.92	&	+60	46	13.4	&	1	&	1	&	0.62	&	-0.79	&	Y64\\
2024	9	12.84336	&	17	19	01.18	&	+61	16	43.2	&	1	&	1	&	0.19	&	-0.17	&	Y65\\
2024	9	12.85477	&	17	18	56.48	&	+61	16	54.7	&	1	&	1	&	0.44	&	0.14	&	Y65\\
2024	9	12.86628	&	17	18	52.04	&	+61	17	03.7	&	1	&	1	&	0.2	&	-0.03	&	Y65\\
2024	9	12.87779	&	17	18	47.64	&	+61	17	11.8	&	1	&	1	&	0.5	&	-0.03	&	Y65\\
2024	9	12.8893	&	17	18	43.44	&	+61	17	17.4	&	1	&	1	&	0.06	&	-0.1	&	Y65\\
2024	9	12.90081	&	17	18	39.48	&	+61	17	21.3	&	1	&	1	&	0.09	&	-0.03	&	Y65\\
2024	9	12.91244	&	17	18	35.66	&	+61	17	23.7	&	1	&	1	&	-0.04	&	-0.22	&	Y65\\
2024	9	12.92782	&	17	18	30.86	&	+61	17	25.8	&	1	&	1	&	-0.78	&	-0.44	&	Y65\\
2024	9	13.8485	&	17	18	05.30	&	+61	47	03.9	&	1	&	1	&	-0.33	&	-0.68	&	Y65\\
2024	9	13.86754	&	17	17	57.79	&	+61	47	18.3	&	1	&	1	&	0.33	&	-0.67	&	Y65\\
2024	9	13.88793	&	17	17	50.3	&	+61	47	28.7	&	1	&	1	&	-0.05	&	0.33	&	Y65\\
2024	9	15.84434	&	17	16	10.55	&	+62	44	45.7	&	1	&	1	&	-0.21	&	0.05	&	Y65\\
2024	9	15.86338	&	17	16	03.04	&	+62	44	58.2	&	1	&	1	&	0.09	&	-0.04	&	Y65\\
2024	9	15.88376	&	17	15	55.46	&	+62	45	06.3	&	1	&	1	&	0.01	&	0.04	&	Y65\\
2024	9	17.84365	&	17	13	58.4	&	+63	38	52.7	&	1	&	1	&	-0.16	&	-0.74	&	Y65\\
2024	9	17.86082	&	17	13	51.66	&	+63	39	01.1	&	1	&	1	&	-0.22	&	0.35	&	Y65\\
2024	9	20.86196	&	17	09	58.04	&	+64	52	53.6	&	1	&	1	&	0.21	&	-0.18	&	Y65\\
2024	9	20.87734	&	17	09	52.26	&	+64	52	55.3	&	1	&	1	&	-0.01	&	-0.04	&	Y65\\
2024	9	20.89048	&	17	09	47.64	&	+64	52	53.7	&	1	&	1	&	0.16	&	-0.02	&	Y65\\
2024	9	21.84305	&	17	08	39.12	&	+65	15	27.1	&	1	&	1	&	0.41	&	0.01	&	Y65\\
2024	9	21.86478	&	17	08	30.58	&	+65	15	30	&	1	&	1	&	0.52	&	0.36	&	Y65\\
2024	9	21.88966	&	17	08	21.65	&	+65	15	26.4	&	1	&	1	&	0.44	&	0.27	&	Y65\\
2024	9	22.84114	&	17	07	10.0	&	+65	37	03.7	&	1	&	1	&	0.05	&	0.14	&	Y65\\
2024	9	22.86519	&	17	07	00.66	&	+65	37	04.4	&	1	&	1	&	0.23	&	0.05	&	Y65\\
2024	9	23.84203	&	17	05	36.94	&	+65	57	41.1	&	1	&	1	&	0.14	&	-0.37	&	Y65\\
2024	9	23.87183	&	17	05	25.56	&	+65	57	39.0	&	1	&	1	&	0.22	&	-0.21	&	Y65\\
2024	9	24.83839	&	17	04	03.45	&	+66	17	22.4	&	1	&	1	&	0.06	&	0.06	&	Y65\\
2024	9	24.85555	&	17	03	56.71	&	+66	17	21.9	&	1	&	1	&	0.2	&	0.14	&	Y65\\
2024	9	24.87203	&	17	03	50.62	&	+66	17	18.2	&	1	&	1	&	-0.15	&	0.19	&	Y65\\
2024	9	25.8363	&	17	02	27.43	&	+66	36	11.3	&	1	&	1	&	0.3	&	-0.58	&	Y65\\
2024	9	25.85233	&	17	02	21.15	&	+66	36	09.6	&	1	&	1	&	-0.18	&	-0.03	&	Y65\\
2024	9	25.86847	&	17	02	15.17	&	+66	36	05.0	&	1	&	1	&	-0.28	&	-0.35	&	Y65\\
2024	9	26.83482	&	17	00	49.93	&	+66	54	10.9	&	1	&	1	&	-0.31	&	-0.34	&	Y65\\
2024	9	26.85011	&	17	00	43.91	&	+66	54	08.8	&	1	&	1	&	-0.39	&	-0.31	&	Y65\\
2024	9	26.8655	&	17	00	38.26	&	+66	54	04.2	&	1	&	1	&	-0.02	&	0.35	&	Y65\\
2024	9	26.88634	&	17	00	31.16	&	+66	53	54.6	&	1	&	1	&	0.38	&	0.4	&	Y65\\
2024	9	26.90718	&	17	00	25.01	&	+66	53	40.9	&	1	&	1	&	0.52	&	0.44	&	Y65\\
2024	9	26.95133	&	17	00	14.74	&	+66	53	06.0	&	1	&	1	&	0.03	&	0.01	&	Y65\\
2024	9	27.24615	&	16	59	59.22	&	+67	01	25.5	&	1	&	1	&	-0.07	&	0.27	&	T15\\
2024	9	27.25283	&	16	59	56.71	&	+67	01	22.3	&	1	&	1	&	-0.01	&	0.31	&	T15\\
2024	9	27.25901	&	16	59	54.45	&	+67	01	19.0	&	1	&	1	&	-0.05	&	0.35	&	T15\\
2024	9	27.83397	&	16	59	11.15	&	+67	11	26.6	&	1	&	1	&	-0.14	&	0.52	&	Y65\\
2024	9	27.83856	&	16	59	09.4	&	+67	11	25.9	&	1	&	1	&	0.05	&	0.32	&	Y66\\
2024	9	27.84556	&	16	59	06.71	&	+67	11	24.4	&	1	&	1	&	-0.25	&	0.13	&	Y64\\
2024	9	27.853	&	16	59	03.82	&	+67	11	21.8	&	1	&	1	&	0.17	&	0.14	&	Y65\\
2024	9	27.86662	&	16	58	58.9	&	+67	11	16.8	&	1	&	1	&	0.06	&	-0.11	&	Y66\\
2024	9	27.8687	&	16	58	58.19	&	+67	11	15.9	&	1	&	1	&	-0.41	&	0.21	&	Y64\\
2024	9	27.87868	&	16	58	54.83	&	+67	11	10.9	&	1	&	1	&	-0.26	&	0.15	&	Y65\\
2024	9	27.89084	&	16	58	50.98	&	+67	11	03.9	&	1	&	1	&	-0.36	&	0.25	&	Y64\\
2024	9	27.89679	&	16	58	49.24	&	+67	10	59.8	&	1	&	1	&	-0.27	&	0.29	&	Y66\\
2024	9	27.91348	&	16	58	44.6	&	+67	10	47.5	&	1	&	1	&	0.29	&	0.26	&	Y64\\
2024	9	28.83442	&	16	57	31.24	&	+67	28	00.7	&	1	&	1	&	-0.08	&	0.52	&	Y65\\
2024	9	28.88381	&	16	57	13.72	&	+67	27	39.0	&	1	&	1	&	-0.11	&	0.19	&	Y65\\
2024	9	28.89529	&	16	57	11.15	&	+67	27	36.2	&	1	&	1	&	0.01	&	0.15	&	950\\
2024	9	28.90177	&	16	57	09.31	&	+67	27	31.6	&	1	&	1	&	0.04	&	0.19	&	950\\
2024	9	28.90825	&	16	57	07.51	&	+67	27	26.9	&	1	&	1	&	0.21	&	-0.13	&	950\\
2024	9	28.91473	&	16	57	05.83	&	+67	27	21.9	&	1	&	1	&	0.02	&	0.06	&	950\\
2024	9	28.92122	&	16	57	04.32	&	+67	27	16.6	&	1	&	1	&	0.27	&	-0.54	&	950\\
2024	9	28.9277	&	16	57	02.74	&	+67	27	11.4	&	1	&	1	&	0.24	&	-0.19	&	950\\
2024	9	29.82906	&	16	55	53.02	&	+67	44	00.0	&	1	&	1	&	0.15	&	-0.32	&	Y65\\
2024	9	29.84847	&	16	55	45.62	&	+67	43	53.9	&	1	&	1	&	0.23	&	0.01	&	Y65\\
2024	10	1.83749	&	16	52	27.89	&	+68	14	20.4	&	1	&	1	&	0.27	&	-0.04	&	Y65\\
2024	10	1.85973	&	16	52	20.02	&	+68	14	08.9	&	1	&	1	&	0.66	&	-0.17	&	Y65\\
2024	10	1.88199	&	16	52	12.98	&	+68	13	54.1	&	1	&	1	&	0.24	&	-0.16	&	Y65\\
2024	10	4.83929	&	16	47	19.41	&	+68	56	55.2	&	1	&	1	&	-0.75	&	0.48	&	Y65\\
2024	10	4.88548	&	16	47	05.0	&	+68	56	22.5	&	1	&	1	&	0.21	&	0.18	&	Y65\\
2024	10	4.93239	&	16	46	54.96	&	+68	55	37.7	&	1	&	1	&	0.27	&	-0.04	&	Y65\\
2024	10	5.84908	&	16	45	31.01	&	+69	10	26.6	&	1	&	1	&	0.66	&	-0.17	&	Y65\\
2024	10	5.88813	&	16	45	19.54	&	+69	09	55.9	&	1	&	1	&	0.24	&	-0.16	&	Y65\\
2024	10	5.92928	&	16	45	11.08	&	+69	09	15.9	&	1	&	1	&	-0.75	&	0.48	&	Y65\\
2024	10	7.88296	&	16	41	45.06	&	+69	36	32.0	&	1	&	1	&	0.21	&	0.18	&	Y65\\
2024	10	7.89237	&	16	41	42.59	&	+69	36	23.9	&	1	&	1	&	0.62	&	-0.09	&	Y65\\
2024	10	8.82832	&	16	40	09.68	&	+69	50	13.4	&	1	&	1	&	-0.11	&	-0.55	&	Y65\\
2024	10	8.84586	&	16	40	03.75	&	+69	50	01.4	&	1	&	1	&	0.44	&	0.34	&	Y65\\
2024	10	8.86173	&	16	39	58.69	&	+69	49	48.6	&	1	&	1	&	0.62	&	-0.09	&	Y65\\
2024	10	9.84523	&	16	38	08.18	&	+70	02	50.4	&	1	&	1	&	-0.11	&	-0.55	&	Y65\\
2024	10	9.88126	&	16	37	57.91	&	+70	02	18.8	&	1	&	1	&	0.44	&	0.34	&	Y65\\
2024	10	9.91815	&	16	37	50.18	&	+70	01	42.8	&	1	&	1	&	0.6	&	-0.53	&	Y65\\
2024	10	10.88056	&	16	35	58.18	&	+70	14	55.4	&	1	&	1	&	-0.34	&	0.19	&	Y65\\
2024	10	10.91746	&	16	35	50.83	&	+70	14	18.0	&	1	&	1	&	-0.31	&	-0.31	&	Y65\\
2024	10	14.84781	&	16	27	15.74	&	+71	03	22.0	&	1	&	1	&	0.1	&	-0.55	&	Y65\\
2024	10	14.89339	&	16	27	04.32	&	+71	02	35.6	&	1	&	1	&	-0.75	&	-0.02	&	Y65\\
2024	10	18.84606	&	16	16	51.1	&	+71	44	58.1	&	1	&	1	&	-0.38	&	-0.23	&	Y65\\
2024	10	18.88508	&	16	16	41.38	&	+71	44	12.4	&	1	&	1	&	0.22	&	0.72	&	Y65\\
2024	10	19.8438	&	16	14	03.17	&	+71	54	01.4	&	1	&	1	&	-0.54	&	-0.11	&	Y66\\
2024	10	22.85193	&	16	05	19.37	&	+72	17	31.3	&	1	&	1	&	-0.02	&	0.21	&	Y65\\
2024	10	22.87034	&	16	05	15.17	&	+72	17	09.0	&	1	&	1	&	-0.3	&	-0.46	&	Y65\\
2024	10	24.84329	&	15	59	28.11	&	+72	31	02.0	&	1	&	1	&	0.62	&	-0.24	&	Y65\\
2024	10	24.85969	&	15	59	24.31	&	+72	30	40.0	&	1	&	1	&	-0.1	&	-0.22	&	Y65\\
\hline
\caption{Columns: (1) UTC observation date at the midpoint of the exposure; (2) right ascension; (3) declination; (4) uncertainty in right ascension; (5) uncertainty in declination; (6) observed-minus-computed residual in the $X$ direction; (7) observed-minus-computed residual in the $Y$ direction; (8) Minor Planet Center Observatory Code}
\label{t:astro}
\end{longtable}
\end{document}